\documentclass[aps,prd,twocolumn,showpacs,floats,floatfix,letterpaper,nofootinbib,superscriptaddress,]{revtex4}

\usepackage{amssymb,amsmath,latexsym,mathrsfs}
\usepackage{graphicx}
\usepackage{epsfig}
\usepackage{multirow}
\usepackage{array}
\usepackage{color}

\newcommand{\bea}{\begin{eqnarray}}
\newcommand{\eea}{\end{eqnarray}}

\newcommand{\neff}{N_{\textrm{eff}}}

\newcommand{\mnu}{{\Sigma}m_{\nu}}

\begin{document}


\title{Relic Neutrinos, thermal axions and cosmology in early 2014}
\author{Elena Giusarma}

\author{Eleonora Di Valentino}
\affiliation{Physics Department and INFN, Universit\`a di Roma ``La Sapienza'', Ple Aldo Moro 2, 00185, Rome, Italy}
\author{Massimiliano Lattanzi}
\affiliation{Dipartimento di Fisica e Science della Terra, Universit\`a di Ferrara and INFN,\\
sezione di Ferrara, Polo Scientifico e Tecnologico - Edficio C Via Saragat, 1, I-44122 Ferrara Italy}
\author{Alessandro Melchiorri}
\affiliation{Physics Department and INFN, Universit\`a di Roma ``La Sapienza'', Ple Aldo Moro 2, 00185, Rome, Italy}
\author{Olga Mena}
\affiliation{IFIC, Universidad de Valencia-CSIC, 46071, Valencia, Spain}
\begin{abstract}
We present up to date cosmological bounds on the sum of active neutrino masses as
well as on extended cosmological scenarios with additional thermal
relics, as thermal axions or sterile neutrino species. Our analyses
consider all the current available cosmological data in the
beginning of year 2014, including the very recent and most precise Baryon
Acoustic Oscillation (BAO) measurements from the Baryon Oscillation
Spectroscopic Survey. In the minimal three active neutrino scenario, we find
$\sum m_\nu < 0.22$~eV at $95\%$~CL from the combination of CMB, BAO and 
Hubble Space Telescope measurements of the Hubble constant. 

A non zero value for the sum of the three active neutrino masses of $\sim 0.3$~eV
is significantly favoured at more than $3$ standard deviations when adding the constraints 
on $\sigma_8$ and $\Omega_m$ from the Planck Cluster catalog on galaxy number counts. 
This preference for non zero thermal relic masses disappears almost completely in both the thermal axion and massive sterile neutrino schemes.

Extra light species contribute to the effective
number of relativistic degrees of freedom, parameterised via $\neff$.
We found that when the recent detection of B mode polarization from the BICEP2 experiment
is considered, an analysis of the combined CMB data in the framework of 
LCDM+r models gives $\neff=4.00\pm0.41$, suggesting the presence of an extra 
relativistic relic at more than $95 \%$ c.l. from CMB-only data.

\end{abstract}

\pacs{98.80.-k 95.85.Sz,  98.70.Vc, 98.80.Cq}

\maketitle

\section{Introduction}
In standard cosmology, hot, thermal relics are identified with the three
light, active neutrino flavours of the Standard Model of elementary
particles. The masses of these three neutrino states have an impact
in the different cosmological observables, see Refs.~\cite{sergio,sergio2} for a
detailed description. Traditionally, the largest
effect caused by neutrino masses on the Cosmic Microwave Background
(CMB) anisotropies, is via the \emph{Early Integrated Sachs
Wolfe effect (ISW)}.  Light active neutrino species may turn
non-relativistic close to the decoupling period, affecting 
the gravitational potentials and leaving a signature which
turns out to be maximal around the first acoustic oscillation peak in
the photon temperature anisotropy spectrum.  

More recently, the Planck
satellite CMB data~\cite{planck}, has opened the window to tackle the neutrino mass via
gravitational lensing measurements: neutrino masses are expected to
leave an imprint on the lensing potential (due to the higher expansion rate) at
scales smaller than the horizon when neutrinos turn non relativistic
states~\cite{lensingnu}.  However, the largest effect of neutrino masses
on the several cosmological observables comes from the suppression of galaxy
clustering at small scales. Neutrinos, being hot thermal relics,
possess large velocity dispersions. Consequently,  the
non-relativistic neutrino overdensities will only cluster at wavelengths larger than their
free streaming scale, reducing  the growth of matter density
fluctuations at small scales, see e.g Refs.~\cite{Reid:2009nq,Hamann:2010pw,dePutter:2012sh,Giusarma:2012ph,Zhao:2012xw,Hinshaw:2012fq,Hou:2012xq,
Sievers:2013wk,Archidiacono:2013lva,Giusarma:2013pmn,Archidiacono:2013fha,Riemer-Sorensen:2013jsa,Hu:2014qma}.  
Non degenerate neutrinos have different free streaming scales and in
principle, with perfect measurements of the matter power spectrum, the
individual values of the neutrino masses could be identified. In
practice, the former is an extremely challenging task. 
Cosmological measurements are, for practical purposes, only sensitive
to the total neutrino mass, i.e. to the sum of the three active neutrino masses.

CMB Measurements from the Planck  satellite, including the lensing likelihood and low-$\ell$ polarization measurements from WMAP
9-year data~\cite{Bennett:2012fp}
 provide a limit on the sum of the three active neutrino masses of
$\sum m_\nu<1.11$~eV at $95\%$~CL.  When a prior on the Hubble constant $H_0$ from
the Hubble Space Telescope~\cite{Riess:2011yx} is added in the
analysis, the constraint is strongly tightened, being $\sum
m_\nu<0.21$~eV at $95\%$~CL, due to the huge existing degeneracy between $H_0$
and $\sum m_\nu$, see Ref.~\cite{Giusarma:2012ph}.
The addition of Baryon Acoustic Oscillation (BAO) measurements from the 
Sloan Digital Sky Survey (SDSS)-II Data Release 7~\cite{dr71,dr72}, from 
the WiggleZ survey~\cite{wigglez}, from the Baryon Acoustic Spectroscopic Survey
(BOSS)~\cite{Dawson:2012va}, one of the four surveys of  SDSS-III~\cite{Eisenstein:2011sa}  Data Release 9~\cite{anderson} and from 6dF~\cite{6df}
to Planck CMB measurements  also significantly improves the
neutrino mass constraints, leading to $\sum m_\nu<0.26$~eV at
$95\%$~CL (see also the recent work of \cite{dePutter:2014hza}).

However, the former bounds are obtained assuming that neutrinos are
the only hot thermal relic component in the universe. The existence of
extra hot relic components, as sterile neutrino species and/or thermal
axions will change the cosmological neutrino mass constraints, see
Refs.~\cite{Hamann:2010bk,Giusarma:2011ex,Giusarma:2011zq,Hamann:2011ge,Giusarma:2012ph,Archidiacono:2013lva,Archidiacono:2013fha,Melchiorri:2007cd,Hannestad:2007dd,Hannestad:2008js,Hannestad:2010yi,Archidiacono:2013cha}.  
Massless, sterile neutrino-like particles, arise naturally in the context of
models which contain a dark radiation sector that decouples from the
Standard Model. A canonical example are asymmetric dark matter models,
in which the extra radiation degrees of freedom are produced by the annihilations of the
thermal dark matter component~\cite{Blennow:2012de}, see also Refs.~\cite{Diamanti:2012tg,Franca:2013zxa} for
extended weakly-interacting massive particle models. 
On the other hand, extra sterile massive, light neutrino species, whose existence is not forbidden by any
fundamental symmetry in nature, may help in resolving the so-called
neutrino oscillation anomalies~\cite{Abazajian:2012ys,Kopp:2013vaa}, see
also Refs.~\cite{Melchiorri:2008gq,Archidiacono:2012ri,Archidiacono:2013xxa,Mirizzi:2013kva,Valentino:2013wha}
for recent results on the preferred sterile neutrino masses and abundances considering both
cosmological and neutrino oscillation constraints. Another candidate
is the thermal axion~\cite{PecceiQuinn}, which constitutes the most elegant solution to the strong CP problem,
i.e. why CP is a respected symmetry of Quantum Chromodynamics (QCD) 
despite the existence of a natural, four dimensional, Lorentz and
gauge invariant operator which badly violates CP. Axions are the Pseudo-
Nambu-Goldstone bosons associated to a new global $U(1)_{PQ}$
symmetry,  which is spontaneously broken at an energy scale $f_a$. 
The axion mass is inversely proportional to the axion coupling constant $f_{a}$ 
\bea
 m_a = \frac{f_\pi m_\pi}{  f_a  } \frac{\sqrt{R}}{1 + R}=
0.6\ {\rm eV}\ \frac{10^7\, {\rm GeV}}{f_a}~,
\label{eq:massaxion}
\eea
where $R=0.553 \pm 0.043 $ is the up-to-down quark masses
ratio and $f_\pi = 93$ MeV is the pion decay constant. 
Axions may be copiously produced in the early universe via thermal or
non-thermal processes, providing therefore, a possible hot relic 
candidate in the thermal case, to be considered together with the
standard relic neutrino background. 

Both extra, sterile neutrino species and axions have an associated free
streaming scale, reducing the growth of matter fluctuations at small scales.
Indeed,  it has been noticed by several authors~\cite{Hamann:2013iba,Wyman:2013lza} that
the inclusion of Planck galaxy cluster number counts data~\cite{Ade:2013lmv} in the
cosmological data analyses, favours a non zero value for the sterile neutrino
mass: the free streaming sterile neutrino nature
will reduce the matter power at small (i.e. cluster) scales but will leave unaffected
the scales probed by the CMB.  A similar tendency for $\sum m_\nu>0$ appears, albeit to a smaller
extent~\cite{Hamann:2013iba}, when considering CFHTLens weak lensing constraints on
the clustering matter amplitude~\cite{Heymans:2013fya}.  

Extra dark radiation or light species as neutrinos and axions will also contribute to the effective
number of relativistic degrees of freedom $\neff$, defined as
\begin{equation}
 \rho_{rad} = \left[1 + \frac{7}{8} \left(\frac{4}{11}\right)^{4/3}\neff\right]\rho_{\gamma} \, ,
\end{equation}
where $\rho_{\gamma}$ is the present energy density of the CMB. 
The canonical value $\neff=3.046$ corresponds to the three active
neutrino contribution. If there are extra light species at the Big Bang
Nucleosynthesis (BBN) epoch, the expansion rate of the universe will
be higher, leading to a higher freeze out temperature for the weak
interactions which translates into a higher primordial helium fraction. The
most recent measurements  of deuterium~\cite{Cooke:2013cba} and helium~\cite{Izotov:2013waa} light
element abundances provide the constraint $\neff=3.50\pm 0.20$~\cite{Cooke:2013cba}.

It is the aim of this paper to analyse the constraints on the three
active neutrino masses, extending the analyses to possible scenarios with
additional hot thermal relics, as sterile neutrino species or axions, 
using the available cosmological data in the beginning of this year 2014.
The data combination used here includes also the recent and most
precise distance BAO constraints to date from the BOSS Data Release 11
(DR11) results~\cite{Anderson:2013vga}, see also Refs.~\cite{Samushia:2013yga,Sanchez:2013tga,Chuang:2013wga}.

The structure of the paper is as follows. Section~\ref{sec:params}
describes the different cosmological scenarios with hot thermal relics
explored here and the data used in our numerical analyses. In
Sec.~\ref{sec:results} we present the current limits using the
available cosmological data in the three active neutrino massive
scenario, and in this same scheme but enlarging the hot relic
component firstly with thermal axions, secondly with additional
dark radiation (which could be represented, for instance, by massless
sterile neutrino
species) and finally, with massive sterile neutrino species. We draw our conclusions in Sec.~\ref{sec:concl}.

\section{Cosmological data analyses}
\label{sec:params}
The baseline scenario we analyse here is light active massive
neutrino scheme with three degenerate massive neutrinos, 
described by the parameters:
\begin{equation}
\label{parameter}
  \{\omega_b,\omega_c, \Theta_s, \tau, n_s, \log[10^{10}A_{s}], \sum m_\nu\}~,
\end{equation}
$\omega_b\equiv\Omega_bh^{2}$ and $\omega_c\equiv\Omega_ch^{2}$  
being the physical baryon and cold dark matter energy densities,
$\Theta_{s}$ the ratio between the sound horizon and the angular
diameter distance at decoupling, $\tau$ is the reionization optical depth,
$n_s$ the scalar spectral index, $A_{s}$ the amplitude of the
primordial spectrum and $\sum m_\nu$ the sum of the masses of the
three active neutrinos in eV.  We then consider simultaneously the
presence of two hot relics, both massive neutrinos and axions,
enlarging the former scenario with one thermal axion of mass $m_a$,
see Appendix~\ref{sec:appdn} for details concerning the calculation of
the axion energy density as a function of the cosmic time. 
The other possibility is the existence of extra dark radiation
species, that we have firstly addressed  by introducing a number of  massless
sterile neutrino-like species, parameterized via $\neff$ (together with the
baseline three massive neutrino total mass $\sum m_\nu$). The extra
additional sterile states, if massive,  may help in resolving the so-called
neutrino oscillation anomalies. Consequently, we also constrain here
simultaneously the $\neff$ massive sterile neutrino scenario and the
sum of the three active neutrino masses $\sum m_\nu$. 
The effective number of massive sterile neutrino species is represented by
$\Delta\neff=\neff- 3.046$, and its mass is $m^\textrm{eff}_s$, which
is related to the physical sterile neutrino mass via the relation:
\begin{equation}
\label{parameter}
m^\textrm{eff}_s= (T_s/T_\nu)^3m_s=(\Delta \neff)^{3/4} m_s~,
\end{equation}
being $T_s$ ($T_\nu$) the current temperature of the sterile (active)
neutrino, and assuming that the sterile states are hot thermal relics
with a phase space distribution similar to the active neutrino one. 

Table \ref{tab:priors} specifies the priors considered on the different cosmological
parameters. For our numerical analyses, we have used the Boltzmann CAMB
code~\cite{camb} and extracted cosmological parameters from current data
using a Monte Carlo Markov Chain (MCMC) analysis based on the publicly
available MCMC package \texttt{cosmomc}~\cite{Lewis:2002ah}. 

In particular, we run chains using the Metropolis-Hastings (MH) algorithm to obtain posterior distributions
for the model parameters, given a certain dataset combination.
The only exception is for the measurements of the power spectrum amplitude (described
in the following section), that are included in our analysis by post-processing 
the MH chains that were previously generated without accounting for these data.
The post-processing is done using the technique of importance sampling;
this technique is very reliable when the posterior distributions obtained
after including new data are centered on the same values as the old distributions,
and becomes on the contrary less reliable the more the new posteriors are 
shifted with respect to the old ones. The reason for this is that in this case one
needs to sample from the low-probability tail of the old distribution, that is poorly
explored by the MH algorithm (unless the chains run for a very long time).
We stress this fact since, as we shall see in the following, the inclusion of the data on the power spectrum 
amplitude shifts the posterior for some of the model parameters.

All the cases under consideration (additional massless species, massive sterile neutrinos, and axions)
can be studied with none or minimal to modifications to the CAMB code. In particular,
the massive sterile and axion cases can be reproduced in the Boltzmann code by means of a
suitable reparameterization and by treating, code-wise, the
additional species as massive neutrinos. This relies on the fact 
that, for an equilibrium distribution function, the evolution equations only depend on the mass over temperature
ratio $m_i/T_i$ and on the total density $\Omega_i$ ($i=\mathrm{a},\,\mathrm{s}$).
The equivalence is perfect for thermal sterile neutrinos, because they have a Fermi-Dirac distribution
function like ordinary neutrinos; instead, this is not the case for thermal axions since they
are described by a Bose-Einstein distribution function. We take into
account here the bosonic nature of axions at the background
level, but not in the perturbation equations. However we argue that the error
that we commit in keeping the Fermi-Dirac distribution function in the perturbation equations 
for axions is negligible given the uncertainties on the model parameters. 

\begin{table}[h!]
\begin{center}
\begin{tabular}{c|c}
\hline\hline
 Parameter & Prior\\
\hline
$\Omega_{b}h^2$ & $0.005 \to 0.1$\\
$\Omega_{c}h^2$ & $0.01 \to 0.99$\\
$\Theta_s$ & $0.5 \to 10$\\
$\tau$ & $0.01 \to 0.8$\\
$n_{s}$ & $0.9 \to 1.1$\\
$\ln{(10^{10} A_{s})}$ & $2.7 \to 4$\\
$\sum m_\nu$  [eV] &  $0.06 \to 3$\\
$m_a$ [eV] &  $0.1 \to 3$\\
$\neff$ &  $0 (3.046) \to 10$\\
$m^\textrm{eff}_s$ [eV] &  $0 \to 3$\\

\hline\hline
\end{tabular}
\caption{Uniform priors for the cosmological parameters considered
  here. In the case of the extra relativistic degrees of freedom $\neff$, the numbers refer
  to the massless (massive) case.}
\label{tab:priors}
\end{center}
\end{table}

\subsection{Cosmological data}
\label{sec:data}
\subsubsection{CMB data }

We consider the data on CMB temperature anisotropies measured 
by the Planck satellite (including information on the lensing potential)
 \cite{Ade:2013ktc,Planck:2013kta,Ade:2013tyw}
combined with 9-year polarization data from WMAP \cite{Bennett:2012fp} 
and with additional temperature data from high-resolution CMB experiments,
namely the Atacama Cosmology Telescope (ACT) \cite{Das:2013zf} and 
the South Pole Telescope (SPT) \cite{Reichardt:2011yv}.
   
The likelihood functions associated to these datasets are estimated 
and combined using the likelihood code
distributed by the Planck collaboration, described in Refs. \cite{Planck:2013kta}  
and \cite{Ade:2013tyw}, and publicly
available at Planck Legacy Archive\footnote{\url{http://pla.esac.esa.int/pla/aio/planckProducts.html}}. 
The Planck TT likelihood is constructed following a hybrid approach:
the high-$\ell$ ($\ell \ge 50$) part is based on 
a pseudo-$C_\ell$ technique and uses power spectra estimated from the detectors 
of the 100, 143 and 217 GHz frequency channels, while the
low-$\ell$ ($\ell\le 49$) part uses a Gibbs sampling-based approach
and combines data from all frequencies from 30 to 353 GHz.
We use Planck TT data up to a maximum multipole number of $\ell_{\rm max}=2500$.
These are supplemented by the low-$\ell$ WMAP 9-year polarization
 likelihood, that includes multipoles up to $\ell=23$ \cite{Bennett:2012fp}.
For what concerns the small-scale observations, we follow the approach 
of the Planck collaboration, as implemented in
their likelihood code, and include the ACT spectra presented in Ref. \cite{Das:2013zf}
 and the SPT spectra presented in Ref. \cite{Reichardt:2011yv}.
In particular, the likelihood uses the ACT $148\times148$ spectra in the range 
$1000<\ell<9440$, the ACT $148\times218$ and $218\times218$
 spectra in the range $1500<\ell<9440$, 
and the SPT 95, 150 and 220 GHz spectra in the range $2000<\ell<10000$,
as described in Sec. 4.1 of Ref. \cite{planck}. The primary purpose of 
considering these subsets of the ACT and SPT data is to improve 
the constraints on the unresolved foregrounds. Finally, we use the information
on the gravitational lensing power spectrum estimated from the trispectrum
of the Planck maps, as implemented in the Planck lensing likelihood described
in Ref. \cite{Ade:2013tyw}.

We shall refer to the combination of all the above-mentioned data as the \emph{CMB} data set.

In our analysis of the CMB dataset, we compute the helium abundance
following the BBN theoretical prediction, in which the helium mass
fraction is a function of $\Omega_b h^2$
and $\neff$ (see the BBN section below) and fix the lensing spectrum
normalization to $A_L=1$. 
 We marginalize over all foregrounds
parameters as described in \cite{planck}.  

\subsubsection{Large scale structure data}
We consider here several large scale structure data sets in different
forms. First of all, we include all the available galaxy survey
measurements in the form of  Baryon Acoustic Oscillation (BAO) data.
As a novelty, we add to the existing BAO data sets (SDSS Data
Release 7~\cite{dr71,dr72}, WiggleZ survey~\cite{wigglez},
6dF~\cite{6df}) the most recent and most accurate BAO
measurements to date, arising from the BOSS Data Release 11 (DR11)
results~\cite{Anderson:2013vga}. Using approximately a sample of one
million galaxies and covering $8500$ squared degrees, the DR11 results
 provide the constraints on the spherically averaged distance $D_V /r_d$~\footnote{The value of the sound horizon $r_d$ used for these values
 is obtained using the Eisenstein \& Hu fitting formula~\cite{Eisenstein:1997ik}.} 
to be $13.42\pm 0.13$ and $8.25\pm 0.16$ at redshifts 
$z=0.57$ and $z=0.32$, respectively. We present results separately for
DR11 BAO measurements, as well as the combination of the former
results with other previous BAO measurements, referring to them as
\emph{DR11} and \emph{BAO}, respectively. 
 
We also exploit here the WiggleZ survey large scale structure
measurements in their full matter power spectrum form~\cite{Parkinson:2012vd}, in order to
quantify the benefits of using \emph{shape} measurements of the matter power 
spectrum versus  \emph{geometrical} BAO information in extended
cosmological scenarios with additional degeneracies among the
different parameters, see the earlier work of Refs.~\cite{Hamann:2010pw,Giusarma:2012ph}
where similar comparisons were performed. This data set is referred as
\textit{WZ}, and whenever it is included, the BAO measurement from the
WiggleZ survey is not considered in the BAO data set.

\subsubsection{Supernova luminosity distance and Hubble constant measurements}
Supernova luminosity distance measurements from the first three years
of the Supernova Legacy Survey~\cite{snls} are included in the hot
thermal  dark matter relic bounds presented here, referring to these
data as \emph{SNLS}. 

Our cosmological data analyses will also address the effect of a gaussian prior on the Hubble constant
$H_0=73.8\pm2.4$ km/s/Mpc, accordingly with the measurements from the
Hubble Space Telescope~\cite{Riess:2011yx}. We refer to this prior as
\emph{HST}. 

\subsubsection{Additional data sets: $\sigma_8$ measurements \label{sec:sigma8}}

Measurements of the galaxy power shear spectra by tomographic weak lensing surveys provide a powerful tool to set constraints on the mass distribution in the universe. The amplitude and the shape of the weak lensing signal are sensitive to the normalization of the power spectrum, the so-called $\sigma_8$ parameter 
(which is the standard deviation of the matter density perturbations in a sphere of radius $8$–Mpc$/h$), as well as to the overall matter energy density of the universe, $\Omega_m$. Using six tomographic redshift bins spanning from $z=0.28$ to $z=1.12$, the CFHTLens survey finds $\sigma_8 (\Omega_m/0.27)^{0.46}=0.774^{+0.032}_{-0.041}$~\cite{Heymans:2013fya}. We shall use this constraint in our analyses, applying this constraint to our Monte Carlo Markov chains. 
 
A strong and independent measurement of the amplitude of the power spectrum arises from the abundance of clusters as a function of the redshift, being the cluster redshift distribution a powerful probe of both $\Omega_m$ and $\sigma_8$. The Planck Sunyaev-Zeldovich (SZ) selected clusters catalog, which consists of  189 galaxy clusters with measured redshift in the X range, is the largest SZ cluster sample to date and has provided the constraint $\sigma_8 (\Omega_m/0.27)^{0.3}=0.782\pm 0.010$~\cite{Ade:2013lmv} via the cluster mass function. We will address as well this constraint in our Monte Carlo Markov chain analyses.
These measurements are included in our analysis by post-processing the
chains that were previously generated without accounting for these data. 

\subsubsection{Big Bang Nucleosynthesis} 
The light elements abundance is also sensitive to several cosmological parameters. The primordial abundance
of deuterium is usually considered as an invaluable \emph{baryometer},
since the higher the baryon abundance $\Omega_b h^2$, the less
deuterium survives. On the other hand, while the mass fraction of
helium-4 $^4He$ ($Y_p$) is rather insensitive to $\Omega_b h^2$, it is
directly related to the expansion rate at the BBN period, which
strongly depends on the effective number of relativistic degrees
of freedom $\neff$. As previously stated, if there are extra light
species at the BBN epoch, the expansion rate of the universe will be
higher, leading to a higher freeze out temperature for the weak
interactions which translates into a higher primordial helium fraction
$Y_p$. Here we exploit the primordial deuterium values from
  Ref.~\cite{fabio} $(D/H)_p = (2.87 \pm 0.22) \times
10^{-5}$ as well as the most recent deuterium measurements $(D/H)_p = (2.53 \pm 0.04) \times
10^{-5}$~\cite{Cooke:2013cba}, to compare the cosmological constraints
obtained with these two diferent primordial deuterium estimates,
including also the measurements of the helium mass fraction $Y_p= 0.254 \pm
0.003$~\cite{Izotov:2013waa}. We shall use the former constraints in
the scenarios in which extra relativistic degrees of freedom are expected to be present at the BBN period.

Notice that  Planck CMB data is also sensitive to the value of $Y_p$ via measurements of the CMB damping
tail (high multipole region), and therefore we use the BBN consistency
option of the MCMC software exploited here,
\texttt{cosmomc}~\cite{Lewis:2002ah}, assuming therefore that the value of the
extra relativistic degrees of freedom remains unchanged between the
BBN and the CMB epochs. Then, given a cosmological
model, the theoretical primordial abundance of helium, which is a function of $\Omega_b h^2$
and $\neff$~\footnote{See for instance the fitting functions provided in
Ref.~\cite{fabio}, extracted from the numerical results of the
PArthENopE BBN code~\cite{parthenope}.} is computed, using
AlterBBN~\cite{alterbbn}, a numerical code devoted to calculate the BBN
abundances within non standard cosmologies. We perform a
similar calculation for the deuterium primordial abundance, and then
fit the theoretical expectations for the deuterium and helium
primordial abundances (previously computed for the CMB data analyses in the latter case) to the measurements quoted above, adding the
resulting likelihood in our MCMC analyses by means of a postprocessing of our chains. 

\subsubsection{Consistency of datasets}

We derive our constraints on model parameters using different combinations of
the datasets described in the previous sections. However, in a few cases there are tensions
between datasets, that we describe in the following. We also briefly assess, at least
qualitatively, the effect on parameters of adding these data.

We use the Planck lensing likelihood in all our analyses. The lensing likelihood is based
on the information encoded in the 4-point correlation function (i.e., the trispectrum) of CMB temperature
anisotropies. On the other hand, lensing also directly affects the CMB power spectrum. 
As explained in Sec. 5.1 of Ref. \cite{planck}, there is a slight tension between the lensing amplitudes 
that are inferred from the trispectrum and from the power spectrum. In particular, while the 
former is consistent with the value expected in $\Lambda$CDM, the temperature
power spectrum shows a mild preference for a higher lensing power. Since the effect of
increasing the neutrino mass is similar to that of a smaller lensing amplitude (as 
both result in a suppression of power at small scales), including the lensing likelihood 
tends to shift the value of the total neutrino mass to larger values  \cite{planck}. 
Instead, the inclusion of the lensing likelihood does not change significantly the constraints
on the effective number of relativistic degrees of freedom, at least for the $\Lambda$CDM 
model. 

Another piece of information that is in tension with the corresponding Planck estimate
is the value of the Hubble constant inferred from astrophysical measurements,
as discussed in Sec. 5.3 of Ref. \cite{planck}. 
This includes the HST value used in our analysis, $H_0=73.8\pm2.4$ km/s/Mpc, that is discrepant
with the Planck $\Lambda$CDM estimate $H_0=67.3\pm1.2$ km/s/Mpc at more than 2$\sigma$,
although it should always be remembered that CMB estimates are highly model dependent.
The reasons for this discrepancy are, to date, not yet well understood and are
a matter of intense debate in the community.
It is however possible that this tension is relieved in some extensions of the
standard $\Lambda$CDM model. For this reason, we have decided to consider 
the HST data in some of our enlarged datasets.

Finally, we use the $\sigma_8$ measurements from the CHFTLens survey
 and from the Planck SZ cluster counts, as reported in Sec. \ref{sec:sigma8}. 
These values are however both discrepant with the value estimated from Planck CMB at 
the 2$\sigma$ level (see discussion in Sec. 5.5 of Ref. \cite{planck}).
This tension has not yet been explained either, but it could be related to the
difficulties in adequately modelling selection biases and calibrating cluster masses.
As in the case of the Hubble constant, however, there is the possibility that the discrepancy
is alleviated in some extended cosmological models (like for example those that include
the neutrino mass as a free parameter). Following the same rationale as for the inclusion of the HST data,
we have derived constraints from enlarged datasets that include the $\sigma_8$ measurements. 
These should however be regarded as quite un-conservative.


\begin{table*}
\begin{center}\footnotesize
\begin{tabular}{lcccccccc}
\hline \hline
                        & CMB+DR11  & CMB+DR11 & CMB+DR11 & CMB+DR11 & CMB+ DR11 & CMB+DR11 & CMB+DR11 & CMB+ DR11 \\
                        &      &              +HST   & +WZ      &  +WZ+HST              & +WZ+BAO+HST & +BAO& +BAO+HST &+BAO+SNLS\\
\hline
\hspace{1mm}\\
               
${\mnu}$ [eV] &  $<0.25$ & $<0.22$  & $<0.25$ & $<0.23$ & $<0.24$  & $<0.26$ & $<0.22$ & $<0.23$\\ 
\hspace{1mm}\\
\hline\\
SZ Clusters \&& & & &  & & & & \\
CFHTLens & & & &  & & & &\\
\hline\\
${\mnu}$ [eV]&  $0.30_{-0.14}^{+0.12}$ & $0.25_{-0.13}^{+0.12}$  & $0.27_{-0.13}^{+0.14}$ & $0.25_{-0.11}^{+0.10}$ & $0.26_{-0.13}^{+0.18}$ & $0.29_{-0.12}^{+0.13}$ & $0.24_{-0.12}^{+0.10}$ & $0.27_{-0.13}^{+0.12}$\\
\hspace{1mm}\\
\hline\\
SZ Clusters & & & &  & & & & \\
\hline\\
${\mnu}$ [eV] &  $0.30_{-0.14}^{+0.12}$ & $0.25_{-0.13}^{+0.13}$   & $0.27_{-0.13}^{+0.12}$ & $0.24_{-0.10}^{+0.10}$ & $0.25_{-0.13}^{+0.17}$\  &$0.29_{-0.12}^{+0.13}$  & $0.23_{-0.12}^{+0.10}$ & $0.27_{-0.13}^{+0.11}$\\ 
\hspace{1mm}\\
\hline\\
 CFHTLens& & & &  & & & &\\
\hline\\
 ${\mnu}$ [eV]  &    $<0.33$  & $<0.28$  & $<0.30$ & $<0.27$ & $<0.28$  & $<0.33$ & $<0.27$ & $<0.30$ \\ 
\hspace{1mm}\\
\hline
\hline
\end{tabular}
\caption{$95\%$~CL constraints on the sum of the neutrino masses, ${\mnu}$, from
  the different combinations of data sets explored here.}
\label{tab:mnu}
\end{center}
\end{table*}

\section{Results} 
\label{sec:results}
\subsection{Massive neutrinos}
\label{sec:st}
We present here the results on our baseline scenario with three active
neutrino degenerate species. Table~\ref{tab:mnu} depicts the $95\%$~CL
constraints on the sum of the three active neutrino masses $\sum m_\nu$. 
Notice that, without the inclusion of the constraints on $\sigma_8$ and $\Omega_m$ the upper limits 
on the neutrino mass are mostly driven by the new BOSS DR11 BAO measurements,  being the tightest limit 
$\sum m_\nu < 0.22$~eV at $95\%$~CL from the combination of CMB data, BAO and HST measurements of the Hubble constant. 
 However, since there exists a well known discrepancy on the measured
 value of $H_0$ from the Planck and the HST experiments~\cite{planck}, we have also considered the combination of 
CMB and BAO data with SNLS Supernovae Ia luminosity distance measurements. Such a combination provides an upper  $95\%$~CL limit of  
$\sum m_\nu < 0.23$~eV, in perfect agreement with the findings of the
recent BOSS results\cite{Sanchez:2013tga} using the full shape of the
clustering correlation function.  
The addition of the constraints on $\sigma_8$ and $\Omega_m$ from the 
CFHTLens survey displaces the bounds on the neutrino mass to higher
values, the reason for that being the lower $\sigma_8$ preferred by
CFHTLens weak lensing measurements. Due the poor constraining power of
the weak lensing data the neutrino mass bounds are not significantly
altered. On the other hand, when adding the constraint on $\sigma_8$ and $\Omega_m$
from the Planck-SZ cluster catalog on galaxy number counts, a non zero
value for the sum of the three active neutrino masses of $\sim 0.3$~eV
is favoured at $4\sigma$. In particular, the combination of CMB
data with BAO measurements from BOSS DR11, WiggleZ power spectrum
(full shape) data and a prior on $H_0$ from HST after considering the
inclusion of Planck SZ clusters information leads to the value $\sum
m_\nu =0.24_{-0.10}^{+0.10}$~eV at $95\%$~CL. The combination of weak
lensing data and galaxy number counts data is mostly driven by the
latter and therefore the constraints do not change significantly with
respect to the case in which the analyses are performed with galaxy
cluster counts information only. A similar effect, although in a slightly different scenario and
different data sets, was found by Refs.~\cite{Hamann:2013iba,Wyman:2013lza}.

\begin{figure*}
\begin{tabular}{c c}
\includegraphics[width=8cm]{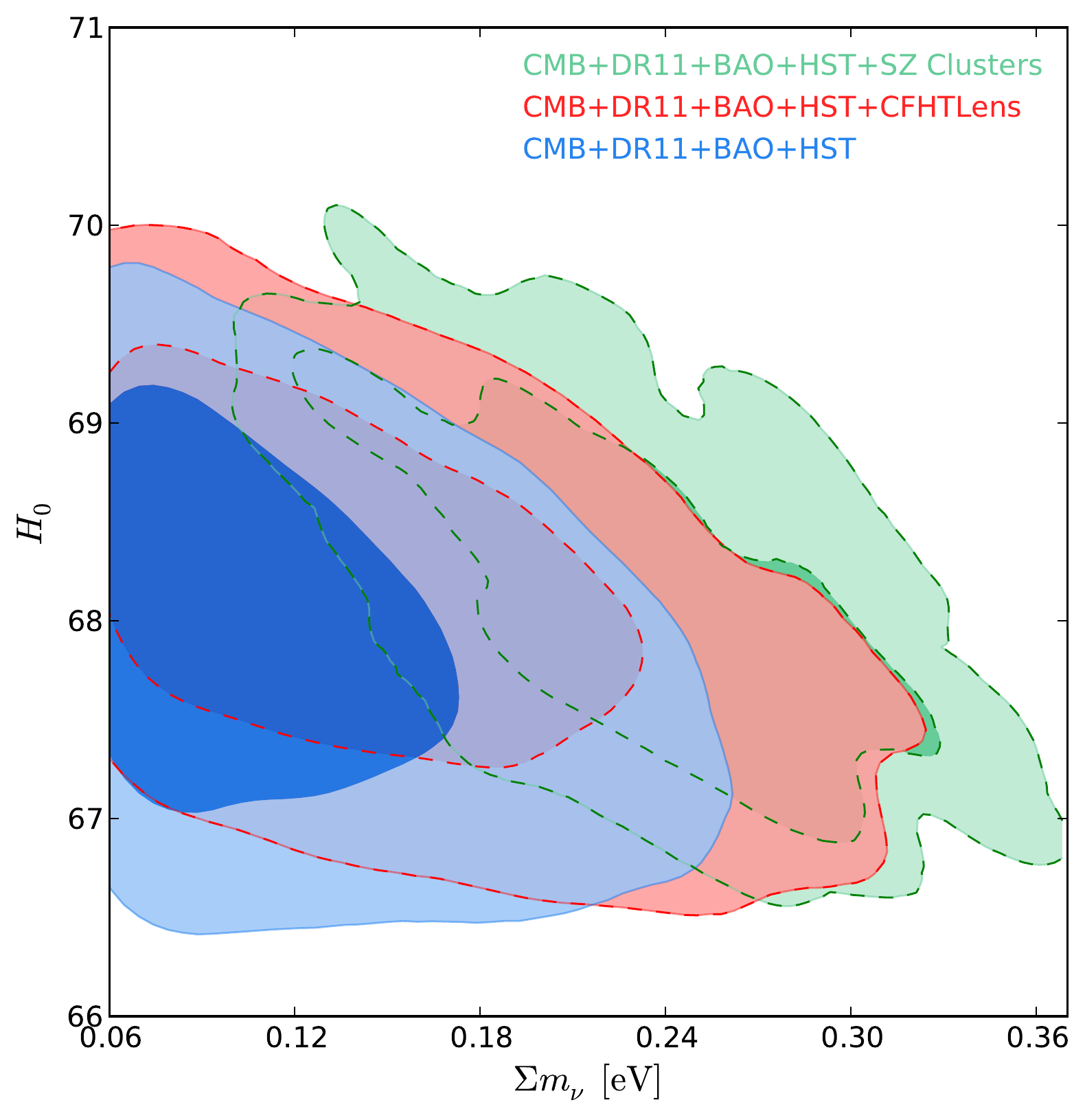}&\includegraphics[width=8.4cm]{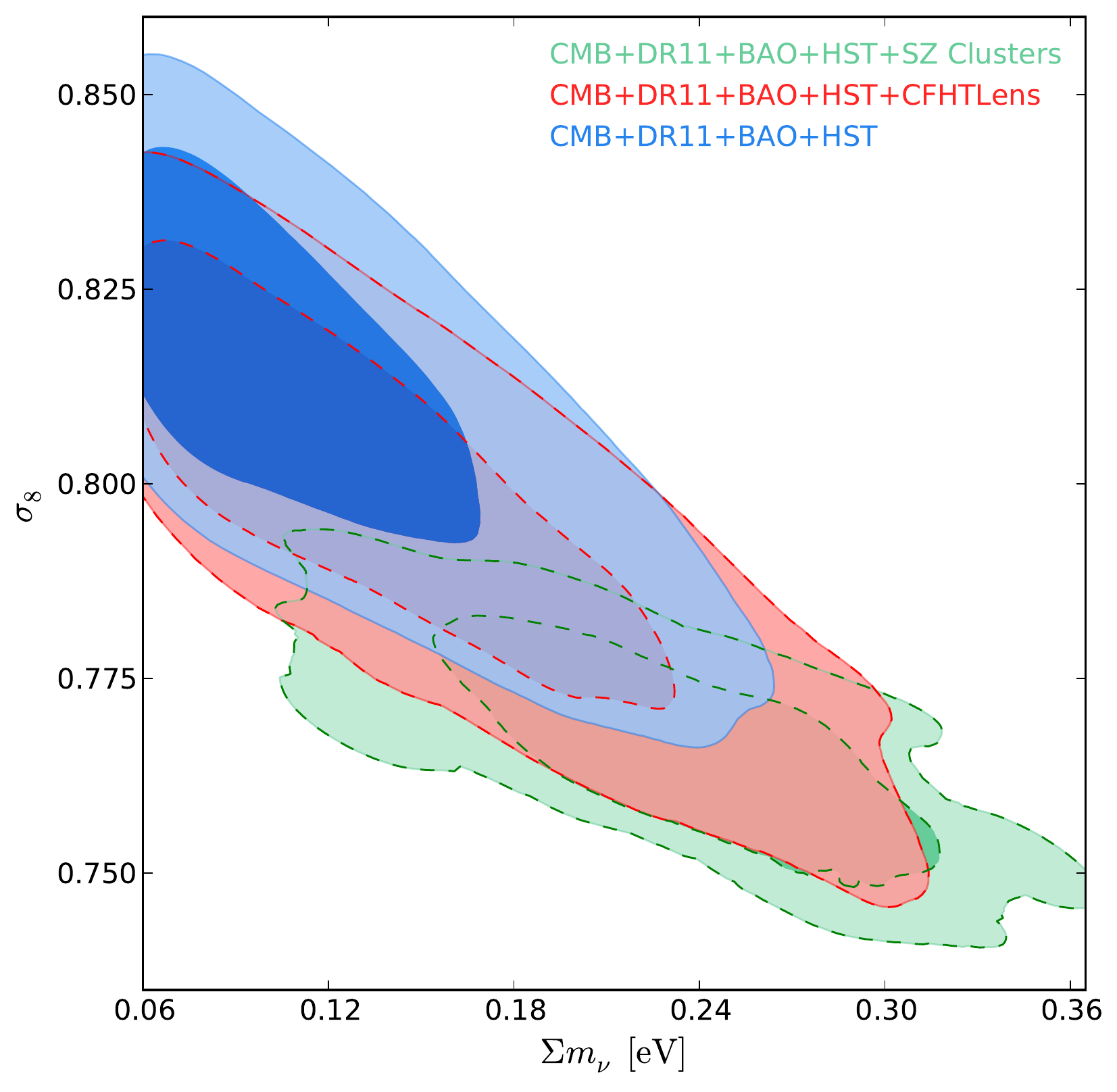}\\
\end{tabular}
 \caption{Left panel: the blue contours show the $68\%$ and $95\%$~CL allowed
  regions from the combination of CMB data, BOSS DR11 BAO measurements,
additional BAO measurements and a prior on the Hubble constant from
HST in the ($\sum m_\nu$ (eV), $H_0$) plane. The red (green) contours depict the results when the $\sigma_8-\Omega_m$  weak lensing
(galaxy number counts) constraint is added in the analysis.
 Right panel: as in the left panel but in the ($\sum m_\nu$ (eV), $\sigma_8$) plane.}
\label{fig:mnu}
\end{figure*}

Figure~\ref{fig:mnu} illustrates our findings for three possible data
combinations: CMB data, combined with BOSS DR11 BAO measurements,
additional BAO measurements and a prior on the Hubble constant from
HST (depicted by the blue contours); and the same data combination but
considering also the $\sigma_8-\Omega_m$  weak lensing
(galaxy number counts) constraint (depicted by the red (green)
contours).  The left panel
depicts the very well known degeneracy in the ($\sum
m_\nu$ (eV), $H_0$) plane, showing the $68\%$ and $95\%$~CL allowed
  contours by the different data sets specified above. Considering CMB
  data only, a higher value of $\sum m_\nu$ can be compensated by a
  decrease on the Hubble constant $H_0$ since the shift induced in the
  distance to the last scattering surface caused by a larger $\sum
  m_\nu$ can be compensated by a lower $H_0$. Notice that when Planck SZ cluster information on the $\sigma_8-\Omega_m$
  relationship is added, the allowed neutrino mass regions are displaced and
  a non zero value for the sum of the three active neutrino masses is
  favoured at $\sim 4\sigma$.  The right panel of Fig.~\ref{fig:mnu} shows the $68\%$ and $95\%$~CL allowed
  regions in the ($\sum m_\nu$ (eV), $\sigma_8$) plane. The allowed contours of
  both $\sigma_8$ and $\sum m_\nu$ are considerably displaced after
  considering Planck clusters data. The power spectrum normalization
  $\sigma_8$ has smaller values when neutrinos are massive (due to the
  neutrino free streaming nature), being precisely these smaller
values of $\sigma_8$ those preferred by galaxy cluster number counts.

\subsection{Massive neutrinos and thermal axions}
\label{sec:st3}
\begin{table*}
\begin{center}\footnotesize
\begin{tabular}{lcccccccc}
\hline \hline
                        & CMB+DR11 & CMB+DR11 & CMB+DR11 & CMB+DR11 & CMB+DR11   & CMB+DR11 & CMB+DR11 & CMB+DR11 \\
                        &      &              +HST   & +WZ      &  +WZ+HST              & +WZ+BAO+HST & +BAO& +BAO+HST &+BAO+SNLS\\
\hline
\hspace{1mm}\\
               
${\mnu}$ [eV] &  $<0.24$ & $<0.21$  & $<0.24$ & $<0.22$ & $<0.21$  & $<0.23$ & $<0.20$ & $<0.22$\\ 
\hspace{1mm}\\
$m_a$ [eV]  &  $<0.79$ &  $<0.77$ &  $<0.65$ &  $<0.62$ &  $<0.59$ &  $<0.74$ &  $<0.75$ &  $<0.76$\\
\hspace{1mm}\\
\hline\\
SZ Clusters \&& & & &  & & & & \\
CFHTLensing & & & &  & & & &\\
\hline\\
 ${\mnu}$ [eV] & $<0.36$ & $<0.27$  & $0.21_{-0.13}^{+013}$ & $<0.32$ & $<0.30$  & $<0.31$ & $<0.28$ &  $<0.31$\\ 
\hspace{1mm}\\
$m_a$ [eV]  &    $<1.08$ &  $<1.09$ & $<0.88$ &  $<0.81$& $<0.77$& $<1.12$ & $0.63_{-0.49}^{+0.47}$&   $0.58_{-0.48}^{+0.50}$\\
\hspace{1mm}\\
\hline\\
SZ Clusters & & & &  & & & & \\
\hline\\
   ${\mnu}$ [eV]   & $<0.36$ & $<0.27$  & $0.20_{-0.14}^{+013}$  & $<0.32$ & $<0.30$  & $<0.31$ & $<0.27$ &  $<0.31$\\
\hspace{1mm}\\
  $m_a$ [eV]      &  $<1.07$ &  $<1.07$ &  $<0.87$&  $<0.81$ &  $<0.77$  &  $<1.10$ & $0.62_{-0.48}^{+0.46}$ & $0.57_{-0.47}^{+0.50}$\\
\hspace{1mm}\\
\hline\\
CFHTLens & & & &  & & & &\\
\hline\\
${\mnu}$ [eV] &    $<0.29$  & $<0.24$  & $<0.28$ & $<0.25$ & $<0.25$  & $<0.27$ & $<0.24$ & $<0.26$  \\ 
\hspace{1mm}\\
$m_a$ [eV]  & $<0.94$ &  $<0.95$ &  $<0.74$ &  $<0.68$ &  $<0.67$ &  $<0.96$ &  $<0.94$ & $<0.98$ \\
\hspace{1mm}\\
\hline\\
BBN & & & &  & & & &\\
\hline\\
${\mnu}$ [eV] $(D/H)_p$\cite{Cooke:2013cba}&  $<0.27$ & $<0.24$  & $<0.26$ & $<0.27$ & $<0.25$  & $<0.27$ & $<0.23$ & $<0.24$\\
\hspace{1mm}\\
${\mnu}$ [eV] $(D/H)_p$\cite{fabio} &  $<0.24$ & $<0.20$  & $<0.23$ & $<0.21$ & $<0.21$  & $<0.22$ & $<0.20$ & $<0.22$\\
\hspace{1mm}\\
$m_a$ [eV]  $(D/H)_p$\cite{Cooke:2013cba}& $<1.15$ &  $<0.76$ & $<0.60$ &  $<0.59$ &  $<0.57$ & $<0.79$ &  $<0.77$ &  $<1.38$  \\
\hspace{1mm}\\
 $m_a$ [eV]  $(D/H)_p$\cite{fabio}&  $<0.82$ & $<0.80$  & $<0.67$ & $<0.64$ & $<0.61$  & $<0.77$ & $<0.78$ & $<0.79$\\
\hspace{1mm}\\
\hline
\hline
\end{tabular}
\caption{$95\%$~CL constraints on the sum of the neutrino masses,
  ${\mnu}$, and on the axion mass, $m_a$, both in eV, from
  the different combinations of data sets explored here. When
   BBN bounds are included, the first (second) raw refers to the constraints
  obtained combining  the primordial deuterium values from
  Ref.~\cite{Cooke:2013cba} (\cite{fabio}) $(D/H)_p = (2.53 \pm 0.04) \times
10^{-5}$ ($(D/H)_p = (2.87 \pm 0.22) \times
10^{-5}$) with measurements of the helium mass fraction $Y_p= 0.254 \pm
0.003$ from Ref.~\cite{Izotov:2013waa}.}
\label{tab:ma}
\end{center}
\end{table*}

In this section we present the constraints on a scenario including
both massive neutrinos and a thermal axion. 
Table~\ref{tab:ma} presents the constraints on the sum of the three
active neutrino masses and on the axion mass (both in eV) for the
different cosmological data combinations considered here. Notice that
BBN bounds are also quoted here since a thermal axion will also
contribute to the extra radiation component at the BBN period, by an
amount given by:
\begin{equation}
\Delta \neff =\frac{ 4}{7}\left(\frac{3}{2}\frac{n_a}{n_\nu}\right)^{4/3}~,
\end{equation}
being $n_a$ the current axion number density  and $n_\nu=112$~
cm$^{-3}$, the current number density of each active neutrino plus
antineutrino flavour. We have applied the BBN consistency
relation in our MCMC analyses of Planck data, to compute the Helium
mass fraction as a function of $\Delta \neff$. Nevertheless the bounds on neutrino and
axion masses are not significantly affected if the Helium mass
fraction is kept fixed for CMB purposes.  Notice that, before applying constraints from
Planck SZ Clusters or CHFTLens constraints on the $\sigma_8-\Omega_m$
relationship, the most stringent $95\%$~CL bounds, without including BBN
bounds, are $\sum m_\nu <0.21$~eV and $m_a<0.59$~eV, considering CMB, BOSS BAO DR11, additional BAO
measurements, WiggleZ power spectrum (full shape) information and the
$H_0$ HST prior. These bounds are in perfect agreement with the
findings of Ref.~\cite{Archidiacono:2013cha}, albeit they are slightly
tighter, mostly due to the more accurate new BOSS BAO measurements.

After considering BBN bounds with deuterium estimates from ~\cite{Cooke:2013cba} (\cite{fabio}) and helium  constraints from Ref.~\cite{Izotov:2013waa}, which constrain the contribution of the thermal axion to
the relativistic degrees of freedom at the BBN epoch, the $95\%$~CL bounds quoted above traslate into $\sum m_\nu <0.25$~eV
and $m_a<0.57$~eV ($\sum m_\nu <0.21$~eV and $m_a<0.61$~eV).

The addition of weak lensing constraints on the $\sigma_8-\Omega_m$ relationship from the CFHTLens
experiment makes the neutrino and axion mass bounds weaker, due to the lower $\sigma_8$ preferred by the
former data set, which favours higher values for the thermal relic
masses. If further information on the $\sigma_8-\Omega_m$ relationship from
the Planck SZ cluster number counts is considered in the MCMC
analyses, there exists evidence for a neutrino mass of $\sim 0.2$~eV 
at the $\sim 3\sigma$ level exclusively for the case in which CMB data
is combined with BOSS BAO DR11 measurements and full-shape power spectrum information from the
WiggleZ  galaxy survey. There exists as well a mild evidence ($\sim
2 \sigma$) for an axion mass of $0.6$~eV for two isolated
cases in which either the HST $H_0$ prior or SNIa luminosity distance measurements are considered in combination
with all the BAO measurements here exploited. However, there is no
evidence for neutrino and axions masses simultaneously.

\begin{figure*}
\begin{tabular}{c c}
\includegraphics[width=8cm]{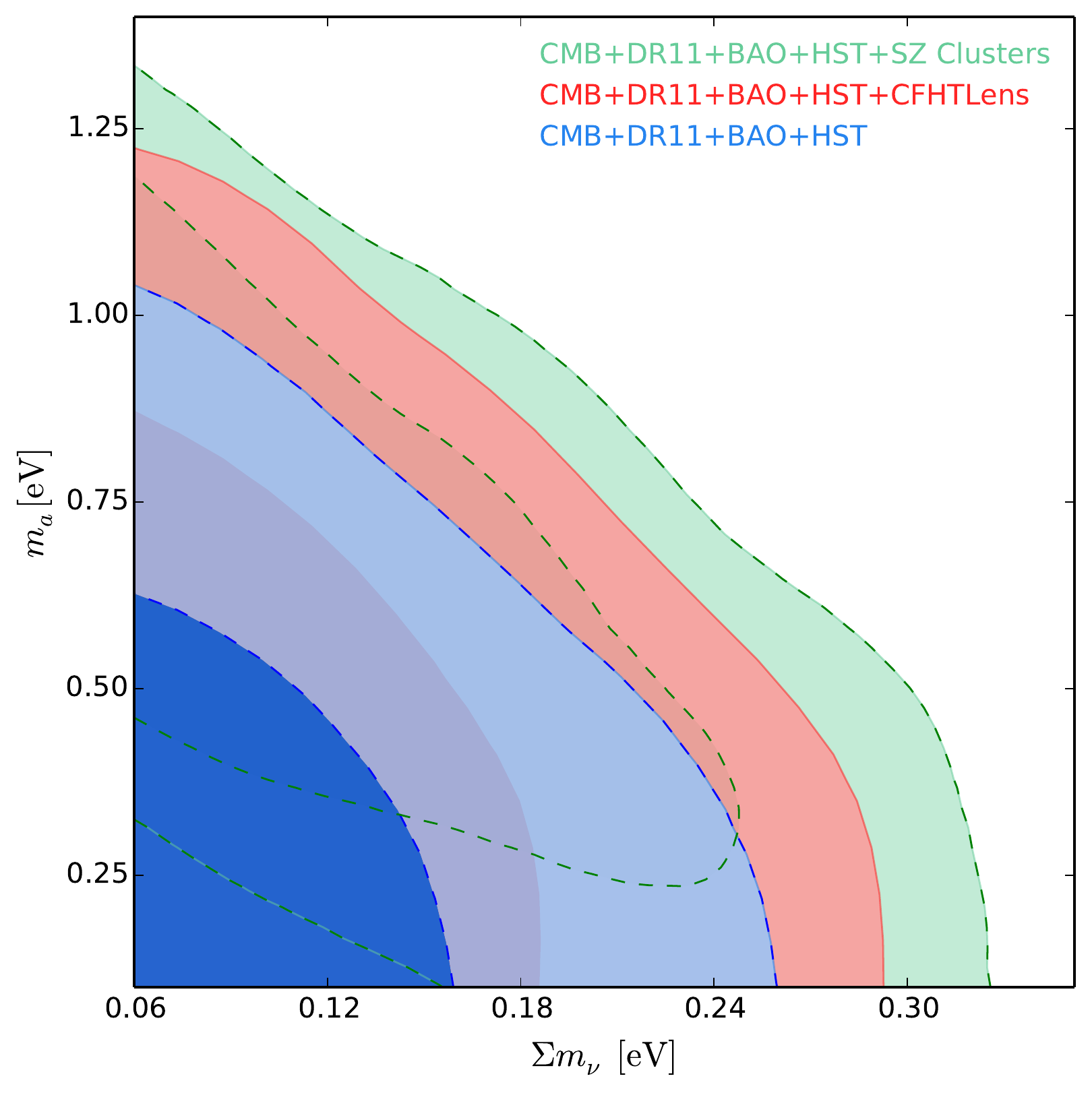}&\includegraphics[width=8.cm]{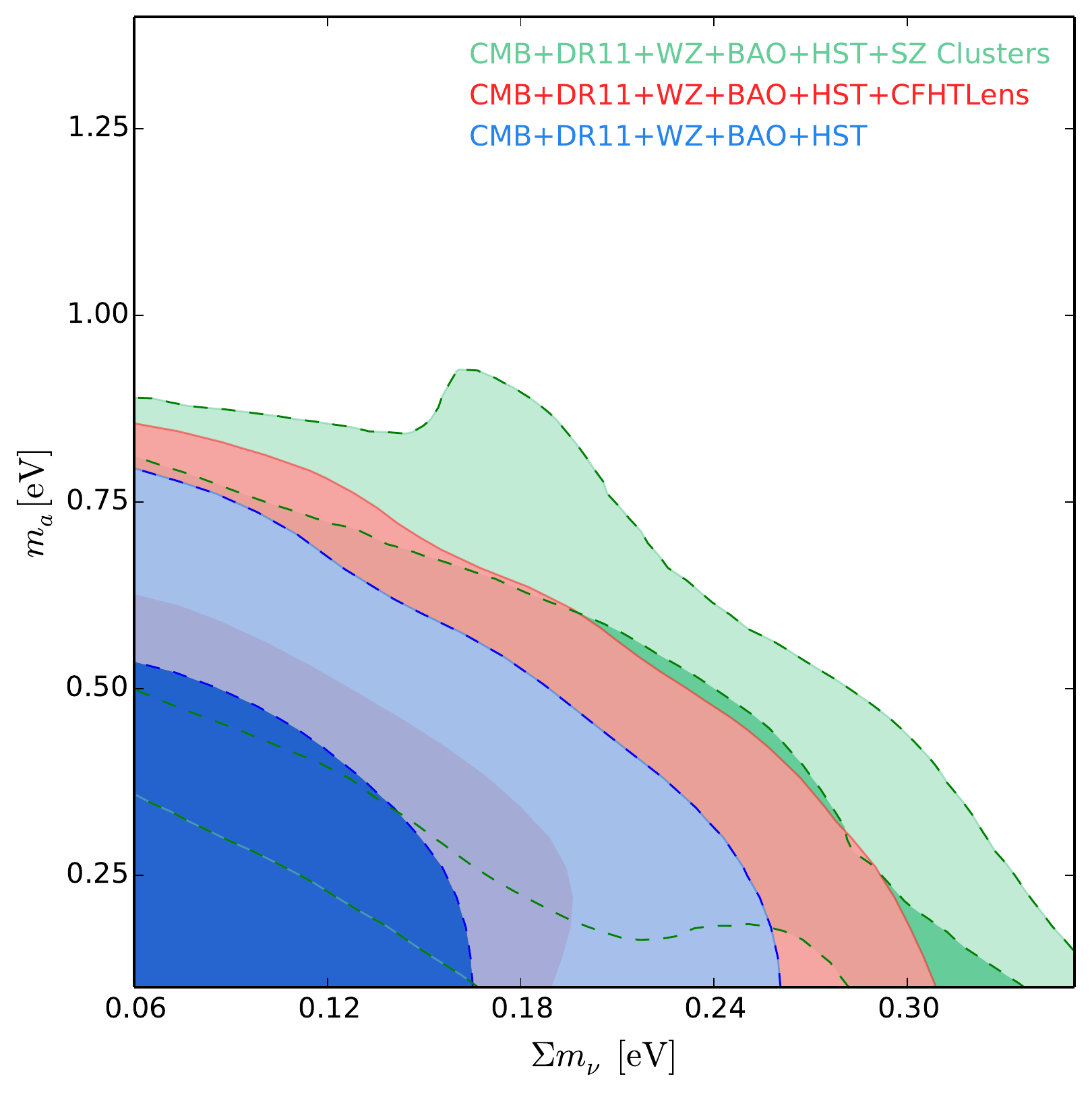}\\
\end{tabular}
 \caption{Left panel: the blue contours show the $68\%$ and $95\%$~CL allowed
  regions from the combination of CMB data, BOSS DR11 BAO measurements,
additional BAO measurements and a prior on the Hubble constant from
HST (depicted by the blue contours) in the ($\sum
m_\nu$ (eV), $m_a$ (eV)) plane. The red (green) contours depict the results when the $\sigma_8-\Omega_m$  weak lensing
(galaxy number counts) constraint is added in the analysis. Right panel: as in the left panel but replacing the WiggleZ BAO geometrical
information by the WiggleZ full-shape matter power spectrum
measurements.}
\label{fig:ma}
\end{figure*}

Figure~\ref{fig:ma}, left panel, depicts the $68\%$ and $95\%$~CL allowed
 regions arising from the combination of CMB data, BOSS DR11 BAO measurements,
additional BAO measurements and a prior on the Hubble constant from
HST in the ($\sum m_\nu$ (eV), $m_a$(eV)) plane. Once the Planck SZ
cluster number counts information on the $\sigma_8-\Omega_m$ relationship is
added, a non zero value of the axion mass is favoured by data at the
$\sim 2.2\sigma$. The right panel of Fig.~\ref{fig:ma} shows the
$68\%$ and $95\%$~CL contours in the ($\sum m_\nu$ (eV), $m_a$(eV))
plane resulting from the analysis of CMB data, BOSS DR11 BAO measurements,
additional BAO measurements - except for the WiggleZ galaxy survey
information which is removed and considered in its full-shape form - 
and the HST $H_0$ prior. Notice that no evidence for non-zero neutrino masses
nor for non-zero axion mass appears in this case.

\subsection{Massive neutrinos and extra dark radiation species}
\label{sec:st2}

\begin{table*}
\begin{center}\footnotesize
\begin{tabular}{lcccccccc}
\hline \hline
                        & CMB+DR11 & CMB+DR11 & CMB+DR11 & CMB+DR11 & CMB+DR11   & CMB+DR11 & CMB+DR11 & CMB+DR11 \\
                        &      &              +HST   & +WZ      &  +WZ+HST              & +WZ+BAO+HST & +BAO& +BAO+HST &+BAO+SNLS\\
\hline
\hspace{1mm}\\
           
${\mnu}$ [eV] &  $<0.31$ & $<0.31$  & $<0.32$ & $<0.34$ & $<0.34$  & $<0.31$ & $<0.31$ & $<0.29$\\ 
\hspace{1mm}\\
$\neff$ &   $3.45_{-0.54}^{+0.59}$ & $3.66_{-0.49}^{+0.52}$ & $3.32_{ -0.62}^{+0.55 }$ & $3.57_{-0.48}^{+0.50}$ & $3.56_{-0.49}^{+0.45}$ &$3.43_{-0.59}^{+0.58}$ &  $3.66_{-0.47}^{+0.48}$ & $3.48_{-0.56}^{+0.58}$ \\
\hspace{1mm}\\
\hline\\
SZ Clusters \&& & & &  & & & & \\
CFHTLensing & & & &  & & & &\\
\hline\\
${\mnu}$ [eV]$\&$ &   $0.37_{-0.18}^{+0.24}$ & $0.37_{-0.20}^{+0.20}$ & $0.32_{ -0.19}^{+0.19 }$ & $0.35_{-0.17}^{+0.16}$ & $0.37_{-0.17}^{+0.26}$ &$0.32_{-0.21}^{+0.18}$ & $0.37_{-0.20}^{+0.18}$ &  $0.32_{-0.17}^{+0.15}$ \\
\hspace{1mm}\\
$\neff$&   $3.32_{-0.55}^{+0.53}$ & $3.54_{-0.54}^{+0.48}$ & $ 3.24_{ -0.70}^{+0.58 }$ & $3.56_{-0.59}^{+0.59}$ & $3.56_{-0.60}^{+1.09}$ & $3.17_{-0.59}^{+0.64}$ & $3.54_{-0.62}^{+60}$ & $3.25_{-0.43}^{+0.47}$ \\
\hspace{1mm}\\
\hline\\
SZ Clusters  & & & &  & & & & \\
\hline\\
${\mnu}$ [eV]&   $0.37_{-0.19}^{+0.24}$ & $0.36_{-0.18}^{+0.18}$ & $0.32_{-0.19}^{+0.19}$ & $0.35_{-0.16}^{+0.17}$ & $0.36_{-0.18}^{+0.26}$&$0.32_{-0.20}^{+0.18}$ & $0.37_{-0.21}^{+0.18}$ &  $0.32_{-0.16}^{+0.15}$ \\ 
\hspace{1mm}\\
$\neff$&  $3.33_{-0.53}^{+0.55}$ & $3.55_{-0.58}^{+0.51}$ & $3.25_{-0.68}^{+0.57}$ & $3.56_{-0.58}^{+0.59}$ & $3.55_{-0.59}^{+0.65}$ & $3.18_{-0.59}^{+0.63}$ & $3.54_{-0.59}^{+0.62}$ & $3.25_{-0.44}^{+0.49}$ \\
\hspace{1mm}\\
\hline\\
 CFHTLens  & & & &  & & & & \\
\hline\\
 ${\mnu}$ [eV]&   $<0.41$ & $<0.44$  & $<0.39$ & $<0.41$ & $<0.42$  & $<0.40$ & $<0.43$ & $<0.39$\\ 
\hspace{1mm}\\
$\neff$& $3.39_{-0.55}^{+0.57}$ & $3.59_{-0.54}^{+0.52}$ & $3.28_{ -0.63}^{+0.58 }$ & $3.55_{-0.47}^{+0.53}$ & $3.54_{-0.47}^{+0.52}$ & $3.33_{-0.61}^{+0.61}$ & $3.58_{-0.50}^{+0.50}$ & $3.37_{-0.55}^{+0.58}$ \\
\hspace{1mm}\\
\hline\\
BBN & & & &  & & & & \\
\hline\\
 ${\mnu}$ [eV] $(D/H)_p$\cite{Cooke:2013cba}&  $<0.27$ & $<0.29$  & $<0.29$ & $<0.24$ & $<0.25$  & $<0.28$ & $<0.32$ & $<0.32$\\ 
\hspace{1mm}\\
 ${\mnu}$ [eV] $(D/H)_p$\cite{fabio} &  $<0.30$ & $<0.28$  & $<0.32$ & $<0.31$ & $<0.32$  & $<0.31$ & $<0.28$ & $<0.28$\\ 
\hspace{1mm}\\
$\neff$ $(D/H)_p$\cite{Cooke:2013cba}& $3.17_{-0.27}^{+0.26}$ & $3.24_{-0.25}^{+0.26}$ & $3.13_{-0.29}^{+0.30}$ & $3.25_{-0.24}^{+0.25}$ & $3.22_{-0.25}^{+0.27}$ & $3.11_{-0.31}^{+0.31}$ & $3.23_{-0.26}^{+0.27}$ & $3.18_{-0.31}^{+0.29}$ \\
\hspace{1mm}\\
 $\neff$ $(D/H)_p$\cite{fabio}& $3.47_{-0.34}^{+0.35}$ & $3.56_{-0.33}^{+0.34}$ & $3.52_{-0.31}^{+0.33}$ & $3.52_{-0.26}^{+0.27}$ & $3.52_{-0.32}^{+0.33}$ & $3.48_{-0.36}^{+0.35}$ & $3.57_{-0.33}^{+0.34}$ & $3.49_{-0.35}^{+0.36}$ \\
\hspace{1mm}\\
\hline
\hline
\end{tabular}
\caption{$95\%$~CL constraints on the sum of the neutrino masses,
  ${\mnu}$, in eV, and on the relativistic degrees of freedom $\neff$ from
  the different combinations of data sets explored here. When
   BBN bounds are included, the first (second) raw refers to the constraints
  obtained combining  the primordial deuterium values from
  Ref.~\cite{Cooke:2013cba} (\cite{fabio}) $(D/H)_p = (2.53 \pm 0.04) \times
10^{-5}$ ($(D/H)_p = (2.87 \pm 0.22) \times
10^{-5}$) with measurements of the helium mass fraction $Y_p= 0.254 \pm
0.003$ from Ref.~\cite{Izotov:2013waa}}.
\label{tab:mnunom}
\end{center}
\end{table*}
We first report here the constraints resulting when considering both
massive neutrinos and $\Delta \neff$ massless dark radiation
species. These massless species may appear in extensions of the
Standard Model of elementary particles containing a dark sector, as,
for instance, in the so-called asymmetric dark matter scenarios. 
In all these models, when the value of $\neff$ is larger than the
canonical 3.046, $\Delta\neff=\neff-3.046$ is related to the extra
density in massless hot relics. On the other hand, if the value of $\neff$ is smaller
than the standard  3.046, the active neutrino temperature is reduced
and there are no extra massless species.

Table~\ref{tab:mnunom} depicts the $95\%$~CL
constraints on the sum of the three active neutrino masses $\sum
m_\nu$ and well as on the total number of dark radiation species $\neff$,
corresponding to the contribution from the three active neutrinos plus
$\Delta \neff$ massless dark radiation species, for the different data
combinations explored here. The bounds on the neutrino mass are less stringent 
than in standard three neutrino massive case due to the 
large degeneracy between $\sum m_\nu$ and $\neff$, since a larger
number of massless sterile neutrino-like species will increase the radiation
content of the universe, and, in order to leave unchanged both the
matter-radiation equality era and the location of the CMB acoustic
peaks, the matter content of the universe must also increase, allowing therefore
for larger neutrino masses. We find $\sum m_\nu < 0.31$~eV and
$\neff=3.45_{-0.54}^{+0.59}$ at $95\%$~CL from the combination of CMB
data and BOSS DR11 BAO measurements. 
When the prior on the value of the Hubble constant from HST  is
included in the analyses, the mean value of $\neff$ and the bound
on the neutrino masses are both mildly larger accordingly to the larger value
of $H_0$ preferred by HST data. The Hubble constant $H_0$
and $\neff$ are positively correlated through measurements of the CMB,
see Ref.~\cite{Hou:2011ec} for a complete description of the effects
of $\neff$ on the CMB . If the value of $\neff$ is increased, in order to
keep fixed both the angular location of the acoustic peaks and the 
matter-radiation equality epoch (to leave unchanged the first peak height
via the early ISW effect), the expansion rate is also increased, implying
therefore a larger $H_0$ and a shorter age of the Universe at recombination.

Since HST measurements point to a higher $H_0$ value, a larger value of $\neff$
will be favoured by data, which also implies a higher neutrino mass
bound due to the strong $\sum m_\nu-\neff$ degeneracy. The $95\%$~CL
constraints from the combination of CMB data, BOSS DR11 BAO
measurements and the HST $H_0$ prior are $\sum m_\nu < 0.34$~eV and
$\neff=3.57_{-0.48}^{+0.45}$. Once the Hubble constant prior from the
HST experiment is added in the analyses, there exists a very mild
preference ($2 \sigma$) for a value of $\neff$ larger than the canonical
expectation of 3.046, agreeing as well with the results of
Ref.~\cite{planck}. 

The addition of the measurements of the deuterium (either from older
estimates~\cite{fabio}, or from the most recent measurements from Ref.~\cite{Cooke:2013cba}) and the helium~\cite{Izotov:2013waa} 
light element abundances, reduce both the mean value and the errors of
$\neff$ significantly.  After the addition of BBN bounds the errors on
 $\neff$ are reduced by a half.  
Table~\ref{tab:mnunom} contains the BBN constraints obtained using the
 fitting functions for the theoretical deuterium and helium
primordial abundances, as a function of $\Omega_b h^2$ and $\neff$, of
Ref.~\cite{fabio} (extracted from the numerical results of the
PArthENopE BBN code~\cite{parthenope}).  We report in the table 
exclusively these constraints because they are
the most conservative ones: we find $\sum m_\nu < 0.24$~eV and
$\neff=3.25_{-0.24}^{+0.25}$ at $95\%$~CL from the analysis of  CMB data,
WiggleZ power spectrum measurements, the HST $H_0$ prior and
BBN light elements abundances information (with the deuterium measurements
from Ref.~\cite{Cooke:2013cba}). Notice that there is no evidence for
$\neff>3$ when considering the most recent estimates of primordial
deuterium abundances. However, if we consider instead previous measurements of
deuterium, as those from Ref.~\cite{fabio},  there exists a $3.5-4\sigma$
preference for $\neff>3$ if HST data is included in the analyses. Without the inclusion of HST data the
preference for $\neff>3$ still persists, albeit at the
$2.5-3\sigma$~CL. As previously stated, the BBN bounds on $\neff$ and $\sum
  m_\nu$ quoted in Tab.~\ref{tab:mnunom} are the most  conservative ones we
  found. Different bounds are obtained if an alternative fitting function
  is used in order to compute the theoretical deuterium and helium
  primordial abundances. We have performed as well such an exercise,
using the fitting functions from
Refs.~\cite{Steigman:2012ve,Cooke:2013cba}
and, in general, the mean value obtained for $\neff$ is larger than
the constraints quoted above. 
 In the case in which recent deuterium measurements are considered in
 the analysis, the mean value of $\neff$ is displaced by $\sim 2
 \sigma$ with respect to the mean values obtained when using 
the fitting function of  \cite{fabio}.  
If previous deuterium measurements~\cite{fabio} are used for our
numerical analyses, the mean value of $\neff$ is also mildly larger
than the mean $\neff$ values obtained when applying the fitting
functions from Ref.~\cite{fabio}. The upper bound on the sum of the 
three active neutrino masses is also larger for the two analyses (with
recent and previous deuterium measurements), 
due to the degeneracy between $\neff$ and $\sum m_\nu$. 
As an example, from the analysis of  CMB data, WiggleZ power spectrum
measurements, the HST $H_0$ prior and BBN light elements abundances
information (with recent deuterium measurements from
Ref.~\cite{Cooke:2013cba}), our analysis point to the following
values: $\neff=3.47^{+0.27}_{-0.27}$ and $\sum m_\nu <0.30$~eV, both
at $95\%$~CL. If previous measurements of deuterium are instead
considered~\cite{fabio}, the $95\%$~CL limits are
$\neff=3.60^{+0.33}_{-0.32}$ and $\sum m_\nu <0.32$~eV. Therefore, a
preference for $\neff > 3$ at the $3.5-4\sigma$ ($2.5-3\sigma$)~CL
with (without) the HST $H_0$ prior included in the analyses
will always be present in the results obtained with the fitting functions
of Refs.~\cite{Steigman:2012ve,Cooke:2013cba}, 
independently of the deuterium measurements exploited.

As in the standard three massive neutrino case, the addition of the
constraints on the $\sigma_8$ and $\Omega_m$ cosmological parameters from the 
CFHTLens survey displaces the bounds on the neutrino mass to higher
values. When adding the $\sigma_8-\Omega_m$ relationship 
from the Planck-SZ cluster catalog on galaxy number counts, a non zero
value for the sum of the three active neutrino masses of $\sim 0.35$~eV
is favoured at $4\sigma$. Notice that in this case the preferred
mean value for $\sum m_\nu$ is higher than in the three massive
neutrino case due to the fact that $\neff$ is a free parameter and
there exists a large degeneracy among $\neff$ and $\sum m_\nu$.
The combination of CMB data with BAO measurements from BOSS DR11, WiggleZ power spectrum
(full shape) data and a prior on $H_0$ from HST after considering the
inclusion of Planck SZ clusters information leads to the values $\sum
m_\nu =0.35_{-0.16}^{+0.17}$~eV and $\neff=3.56_{-0.58}^{+0.59}$ at $95\%$~CL. 

The bounds quoted above have been obtained using the BBN
theoretical prediction for helium in the CMB data analysis. However, it is also
possible  to fix the helium fraction $Y_p$ in the Monte Carlo Markov Chain analyses of
CMB data and assume that $Y_p$ is an independent parameter
constrained by BBN observations only. We have also performed such an
exercise, fixing $Y_p=0.24$, and we find, in general, larger values
 for both the mean value of $\neff$ and its errors, and, consequently,
 a slightly larger bound on the neutrino mass, due to the $\sum
 m_\nu-\neff$ degeneracy. In particular, we find $\sum m_\nu < 0.32$~eV and
$\neff=3.60_{-0.65}^{+0.67}$ at $95\%$~CL from the combination of CMB
data and BOSS DR11 BAO measurements, and $\sum m_\nu < 0.34$~eV and
$\neff=3.84_{-0.56}^{+0.60}$  at $95\%$~CL if a prior from HST on the Hubble
constant $H_0$ is added  to the former data combination. These
findings agree with the results of Ref.~\cite{Archidiacono:2013fha},
where it is also found that the BBN consistency relation leads to a constraint on $\neff$
 closer to the canonical value of $3.046$ than in the case of
fixing $Y_p=0.24$.  Once BBN measurements are considered in the data
analyses, the differences between the analyses with and without the
BBN consistency relation included become irrelevant.

\begin{figure*}
\begin{tabular}{c c}
\includegraphics[width=8cm]{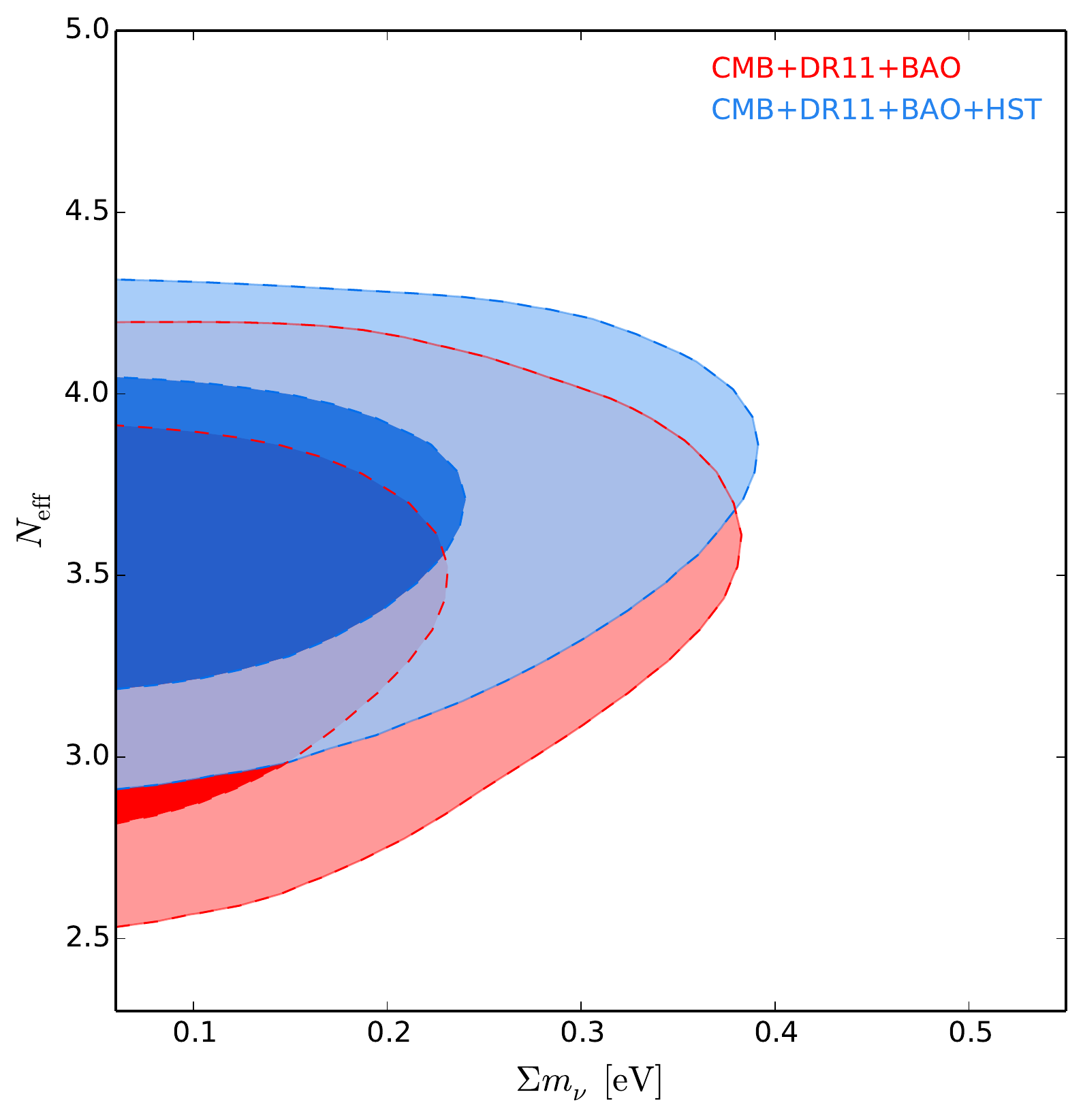}&\includegraphics[width=8.15cm]{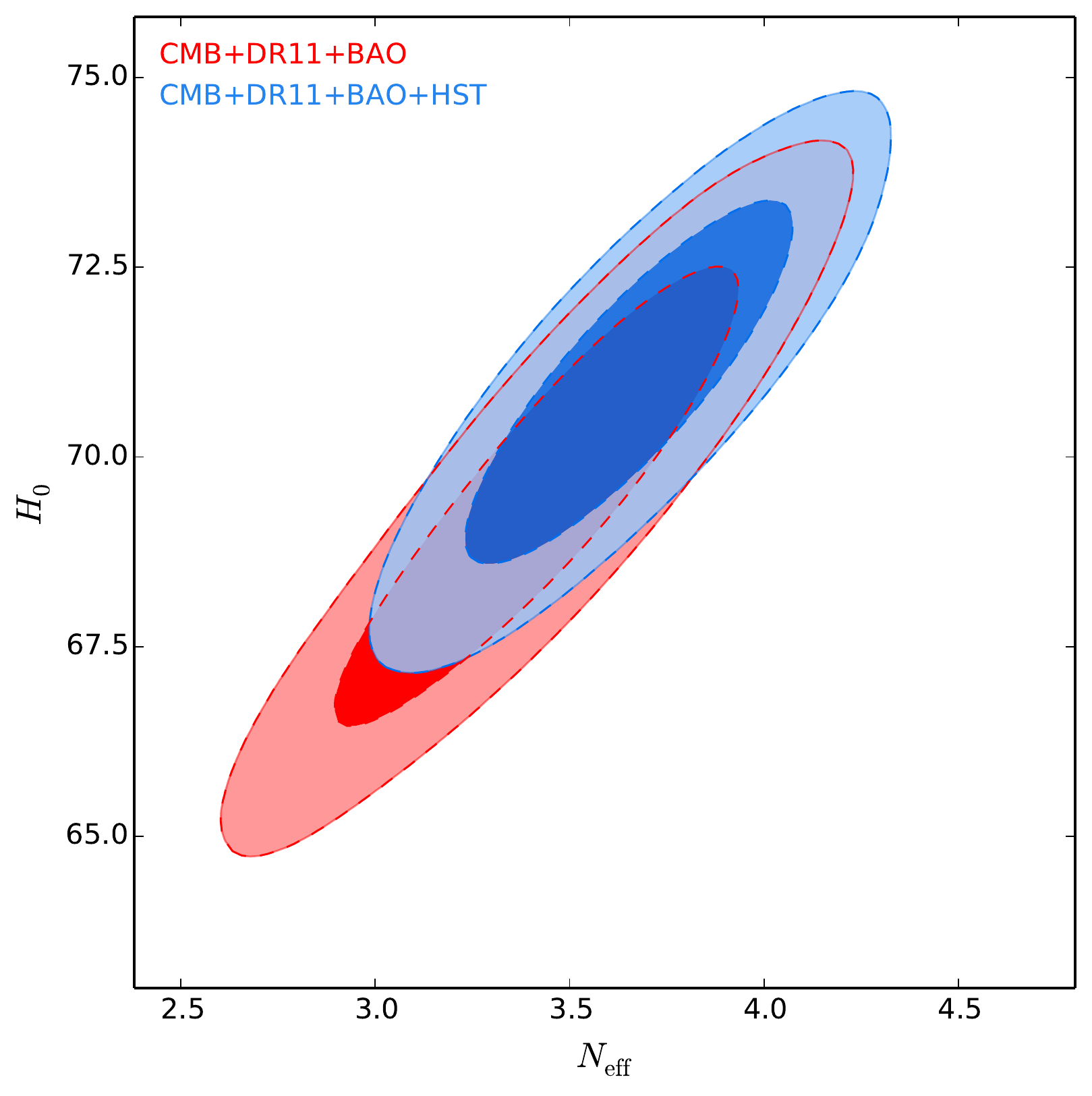}\\
\end{tabular}
 \caption{Left panel: the red contours show the $68\%$ and $95\%$~CL allowed
  regions from the combination of CMB data, BOSS DR11 BAO measurements
  and additional BAO measurements  in the ($\sum
m_\nu$ (eV), $\neff$) plane. The blue contours depict the
  constraints after a prior on the Hubble constant from
HST is added in the analysis. Right panel: as in the left panel but in the ($\neff$, $H_0$) plane.}
\label{fig:nnu}
\end{figure*}

Figure~\ref{fig:nnu}, left panel, shows the degeneracy between 
the $\sum m_\nu$ and the total number of dark radiation species $\neff$ 
(which accounts for the contribution of the three active neutrino
species plus $\Delta \neff$ massless sterile neutrino-like species). The red
contours depict the $68\%$ and $95\%$~CL allowed regions resulting
from the combination of CMB, BOSS DR11 BAO measurements, and previous
BAO measurements. As the value of $\neff$ increases, a larger neutrino
mass is allowed, to leave unchanged both the matter radiation equality
era and  the angular location of the acoustic peaks, as well as the
high of the first acoustic peak via the early ISW effect. 
The blue region denotes the results considering the HST $H_0$ prior as
well in the analysis: notice that the allowed regions are shifted
towards higher values  of $\neff$.   Figure~\ref{fig:nnu}, right panel, illustrates the degeneracy between
$\neff$ and the Hubble constant $H_0$. The color coding is identical
to the one used in the figure shown in the left panel, in which the red contours are related to the $68\%$ and $95\%$~CL allowed
  regions from the combination of CMB data, BOSS DR11 BAO measurements
  and additional BAO measurements and the blue regions refer to the
  constraints after adding a prior on the Hubble constant from
the HST experiment.

\subsection{Massive neutrinos and extra massive sterile neutrino species}
The latest possibility for thermal relics explored in this study is
the case in which there exists three active light massive neutrinos
plus one  massive sterile neutrino species characterised by an effective mass
$m^\textrm{eff}_s$, which reads
\begin{equation}
\label{parameter}
m^\textrm{eff}_s= (T_s/T_\nu)^3m_s=(\Delta \neff)^{3/4} m_s~,
\end{equation}
being $T_s$ ($T_\nu$) the current temperature of the sterile (active)
neutrino species, $\Delta \neff \equiv \neff-3.046=(T_s/T_\nu)^3$ the effective number of degrees of freedom associated to the sterile, 
and $m_s$ its real mass.

\begin{table*}
\begin{center}\footnotesize
\begin{tabular}{lcccccccc}
\hline \hline
                        & CMB+DR11 & CMB+DR11 & CMB+DR11 & CMB+DR11 & CMB+DR11   & CMB+DR11 & CMB+DR11 & CMB+DR11 \\
                        &      &              +HST   & +WZ      &  +WZ+HST              & +WZ+BAO+HST & +BAO& +BAO+HST &+BAO+SNLS\\
\hline\\
               
${\mnu}$ [eV] &  $<0.28$ & $<0.27$  & $<0.28$ & $<0.30$ & $<0.31$  & $<0.30$\,\footnotemark[1] & $<0.29$  & $<0.26$\\ 
\hspace{1mm}\\
$m^\textrm{eff}_s$ [eV] &  $<0.29$ &  $<0.28$ &  $<0.60$ &  $<0.28$ &  $<0.25$ & $<0.27$\,\footnotemark[1]
 & $<0.28$ &  $<0.31$\\
\hspace{1mm}\\
$\neff$ &   $<4.01$ & $3.73_{-0.51}^{+0.51}$ & $<3.89$ & $<4.06$ & $3.64_{-0.48}^{+0.48}$ & $3.57_{-0.50}^{+0.50}$ \footnotemark[1] & $<4.16$ & $<4.02$ \\
\hspace{1mm}\\
\hline\\
SZ Clusters\&& & & &  & & & & \\
CFHTLens & & & &  & & & &\\
\hline\\
${\mnu}$ [eV]  &     $<0.40$ & $<0.43$  & $<0.36$ & $<0.41$ & $<0.43$  & $<0.43$ & $<0.39$  & $<0.37$\\ 
\hspace{1mm}\\
$m^\textrm{eff}_s$ [eV] &   $<0.50$ & $<0.48$  & $<1.37$ & $<0.39$ & $<0.34$  & $<0.49$ & $<0.59$  & $<0.59$\\ 
\hspace{1mm}\\
$\neff$ & $<3.90$ & $3.67_{-0.55}^{+0.49}$ & $<3.77$ & $<4.08$ & $3.67_{-0.45}^{+0.51}$ & $3.47_{-0.39}^{+0.51}$ & $<4.01$ & $<3.85$ \\
\hspace{1mm}\\
\hline\\
SZ Clusters & & & &  & & & & \\
\hline\\
 ${\mnu}$ [eV] &     $<0.40$ & $<0.42$  & $<0.36$ & $<0.41$ & $<0.42$  & $<0.41$ & $<0.39$  & $<0.38$\\ 
\hspace{1mm}\\
$m^\textrm{eff}_s$ [eV] &    $<0.49$ & $<0.48$  & $<1.36$ & $<0.39$ & $<0.34$  & $<0.49$ & $<0.53$  & $<0.59$\\ 
\hspace{1mm}\\
$\neff$ &  $<3.90$ & $3.66_{-0.55}^{+0.49}$ & $<3.77$ & $<4.06$ & $3.66_{-0.45}^{+0.50}$ & $3.46_{-0.38}^{+0.41}$ & $<4.02$ & $<3.85$ \\
\hspace{1mm}\\
\hline\\
CFHTLens & & & &  & & & & \\
\hline\\
${\mnu}$ [eV] &   $<0.35$ & $<0.33$  & $<0.32$ & $<0.35$ & $<0.35$  & $<0.39$ & $<0.34$  & $<0.31$\\ 
\hspace{1mm}\\
 $m^\textrm{eff}_s$ [eV] &$<0.39$ & $<0.39$  & $<1.16$ & $<0.34$ & $<0.29$  & $<0.37$ & $<0.43$  & $<0.47$\\ 
\hspace{1mm}\\
$\neff$ &   $<3.94$ & $3.68_{-0.51}^{+0.51}$ & $<3.85$ & $<4.06$ & $3.63_{-0.49}^{+0.49}$ & $3.50_{-0.44}^{+0.48}$ & $<4.09$ & $<3.94$ \\
\hspace{1mm}\\
\hline\\
BBN & & & &  & & & & \\
\hline\\
 ${\mnu}$ [eV] $(D/H)_p$\cite{Cooke:2013cba}&$<0.28$ & $<0.23$  & $<0.25$ & $<0.24$ & $<0.27$  & $<0.39$ & $<0.24$  & $<0.23$\\ 
\hspace{1mm}\\
   ${\mnu}$ [eV] $(D/H)_p$\cite{fabio}                     &$<0.27$ & $<0.25$  & $<0.28$ & $<0.28$ & $<0.28$  & $<0.29$ & $<0.26$  & $<0.25$\\ 
\hspace{1mm}\\
$m^\textrm{eff}_s$ [eV] $(D/H)_p$\cite{Cooke:2013cba}&  $<0.45$ & $<0.34$  & $<0.37$ & $<0.46$ & $<0.14$  & $<0.24$ & $<0.56$  & $<0.62$\\ 
\hspace{1mm}\\
         $m^\textrm{eff}_s$ [eV] $(D/H)_p$\cite{fabio} &$<0.27$ & $<0.25$  & $<0.29$ & $<0.24$ & $<0.23$  & $<0.27$ & $<0.26$  & $<0.27$\\ 
\hspace{1mm}\\ 
$\neff$ $(D/H)_p$\cite{Cooke:2013cba}& $<3.41$ & $<3.53$ & $<3.49$ & $<3.58$ & $3.28_{-0.21}^{+0.22}$ & $3.25_{-0.17}^{+0.17}$ & $<3.47$ & $<3.43$ \\
\hspace{1mm}\\
$\neff$ $(D/H)_p$\cite{fabio}& $3.48_{-0.35}^{+0.37}$ & $3.59_{-0.34}^{+0.35}$ & $3.45_{-0.38}^{+0.33}$ & $3.56_{-0.34}^{+0.34}$ & $3.56_{-0.32}^{+0.33}$ & $3.50_{-0.36}^{+0.35}$ & $3.59_{-0.45}^{+0.35}$ & $3.50_{-0.37}^{+0.36}$ \\
\hspace{1mm}\\
\hline
\end{tabular}
\footnotetext[1]{These limits have been obtained by imposing an additional prior on the thermal velocity of sterile neutrinos. See discussion in the text for further details.}
\caption{$95\%$~CL constraints on the active (sterile) neutrino masses,
  ${\mnu}$ ($m^\textrm{eff}_s$), in eV, and on the total number of
  massive neutrino species, $\neff$, from
  the different combinations of data sets explored here. When
   BBN bounds are included, the first (second) raw refers to the constraints
  obtained combining  the primordial deuterium values from
  Ref.~\cite{Cooke:2013cba} (\cite{fabio}) $(D/H)_p = (2.53 \pm 0.04) \times
10^{-5}$ ($(D/H)_p = (2.87 \pm 0.22) \times
10^{-5}$) with measurements of the helium mass fraction $Y_p= 0.254 \pm
0.003$ from Ref.~\cite{Izotov:2013waa}.}
\label{tab:mnumeff}
\end{center}
\end{table*}

Table~\ref{tab:mnumeff} depicts the $95\%$~CL
constraints on the active and sterile neutrino masses as well as on
the total number of massive neutrinos $\neff$.  Notice that the mean
value of $\neff$ is, in general, slightly larger than in the case in
which the sterile neutrinos are considered as massless particles due to
the fact that $m^\textrm{eff}_s$  and $\neff$ are positively
correlated. Indeed, there exists a physical lower prior for $\neff$ of 3.046
which is not needed in the case of three active neutrinos plus extra
massless species. We quote exclusively the $95\%$~CL upper limit for the cases
in which the $95\%$~CL lower limit is set by the physical prior of 3.046.
 Concerning the bounds on the sum of the three active
neutrinos, they are more stringent than in the massless sterile
neutrino-like scenario because $\sum m_\nu$ and $m^\textrm{eff}_s$ are
also positively correlated.  As in the massless sterile neutrino-like analyses, larger values of $\neff$
will be favoured by data when HST measurements
are included.  The addition of BBN bounds reduce the errors on
$\neff$ significantly, alleviating the degeneracies between $\neff$
and the active/sterile neutrino masses. 
Table~\ref{tab:mnumeff} contains the BBN constraints obtained using the
 fitting functions for the theoretical deuterium and helium
primordial abundances from Ref.~\cite{fabio}, which, as in the
massless extra dark radiation case, are found to provide the most
conservative bounds. We find $\sum m_\nu < 0.27$~eV,  $m^\textrm{eff}_s<0.14$~eV and
$\neff=3.28_{-0.21}^{+0.22}$ at $95\%$~CL from the analysis of  CMB data,
BOSS DR11 BAO, additional BAO measurements, WiggleZ full-shape large
scale structure information, the HST $H_0$ prior and
BBN light elements abundances information with  the most recent
measurements of the primordial deuterium abundances from
  Ref.~\cite{Cooke:2013cba}, indicating no significant preference for
  $\neff>3$. However, when considering primordial deuterium
  measurements from Ref.~\cite{fabio}, there exists a preference for
  $\neff>3$ at the $3\sigma$ level (mildly stronger when HST data is
  also considered in the analyses). This preference is similar to that found in the extra massless case, although notice that in
this case there exists a lower prior on $\neff=3.046$ and therefore
the mean value of $\neff$ will always be larger than its standard
prediction. If we instead use the theoretical functions for the helium
and deuterium abundances from
Refs.~\cite{Steigman:2012ve,Cooke:2013cba}, we get similar conclusions
to those found in the massless dark radiation case:  a $3-4\sigma$ preference for
  $\neff>3$ is always present. The bounds on the neutrino masses
  are, as in the massless case, mildly loosened. The constraints quoted
  above translate into $\sum m_\nu < 0.28$~eV,  $m^\textrm{eff}_s<0.22$~eV and
$\neff=3.50_{-0.28}^{+0.27}$ ($\sum m_\nu < 0.30$~eV,  $m^\textrm{eff}_s<0.24$~eV and
$\neff=3.64_{-0.33}^{+0.33}$) at $95\%$~CL from the analysis of  CMB data,
BOSS DR11 BAO, additional BAO measurements, WiggleZ full-shape large
scale structure information, the HST $H_0$ prior and
BBN light elements abundances information with  the most recent
measurements of the primordial deuterium abundances from
  Ref.~\cite{Cooke:2013cba} (\cite{fabio}).
  
We have also found that the posterior distribution obtained from the CMB+DR11+BAO dataset
(without the addition of any BBN or $\sigma_8$ information) is multimodal. In fact,
we find that the probability density is significantly different from zero, other than for
 $m_\mathrm{eff} \lesssim 0.3$ eV (as for the other datasates), 
also for $m_\mathrm{eff} \gtrsim 1$ eV. A further inspection of the chains has
shown that these two regions roughly corresponds to the two cases of a hot/warm 
(at recombination) sterile neutrino, with a mass-to-temperature ratio at that time $m_s/T_{s,\mathrm{rec}}
\lesssim 10$, and of a cold sterile with $m_s/T_{s,\mathrm{rec}} \gtrsim 100$. The limits quoted in Tab. V
for the CMB+DR11+BAO dataset, in the basic case where no other information is considered, have
been obtained by postprocessing the chains in order to keep only those models with $m_s/T_{s,\mathrm{rec}}
\lesssim 10$. This is consistent with the purpose of the paper of costraining the presence of
a hot component in addition to active neutrinos. We have also verified that these limits 
are reasonably stable with respect to the choice 
of the value of the mass-to-temperature ratio at which to cut the distribution, as long as this value lies
inside the low-probability region $10 \lesssim m_s/T_{s,\mathrm{rec}} \lesssim 100$. 
It still has to be clarified which, if any in particular, of the BAO datasets is responsible for the appearance
of the ``cold sterile'' region in the posterior probability, and to which feature in the data this is possibly related.
A very preliminar analysis, performed using only one at a time among the DR7, 6dF and WiggleZ BAO datasets, 
seems to show that this effect is mainly driven by the first two datasets, while using only the WiggleZ BAO measurements
yields naturally an upper limit for $m_\mathrm{eff}$ of about $0.3$ eV, without any need to exclude a priori the cold region. However
a more robust and precise assessment of the role of the different datasets would certainly require a more detailed
analysis that goes beyond the scope of the present paper.

Contrarily to the massless dark radiation case (and similarly to the thermal axion scenario), the addition of the
constraints on the $\sigma_8$ and $\Omega_m$ cosmological parameters
from the Planck-SZ cluster catalog on galaxy number counts does not
lead to a non zero value for the neutrino masses. However, the bounds on the
neutrino masses are less stringent when adding the Planck-SZ or the
CFHTLens constraints on the $\sigma_8$ and $\Omega_m$ cosmological
parameters,  due to the lower $\sigma_8$ preferred by the
former data sets, which favours higher values for the thermal relic
masses. After considering the inclusion of Planck SZ clusters and
CFHTLens information to CMB data,
BOSS DR11 BAO and additional BAO measurements and the HST $H_0$ prior, 
the $95\%$~CL bounds on the active and the sterile neutrino parameters are $\sum m_\nu < 0.39$~eV,  $m^\textrm{eff}_s<0.59$~eV and
$\neff<4.01$.

The bounds quoted in Tab.~\ref{tab:mnumeff} have been obtained using the BBN theoretical prediction for
helium in the CMB data analysis, as in the case of extra massless
species. We have also performed in this massive case the exercise of
fixing the helium fraction $Y_p$ in the Monte Carlo Markov Chain analyses of
CMB data and assume that $Y_p$ is an independent parameter
constrained by BBN observations only. Again, as in the massless case,
we find larger values for the mean value of $\neff$ (and,
consequently, slightly larger bounds on both the active and sterile neutrino
 masses) when neglecting the BBN consistency relation in the MCMC
 analyses.

\begin{figure*}
\begin{tabular}{c c}
\includegraphics[width=8cm]{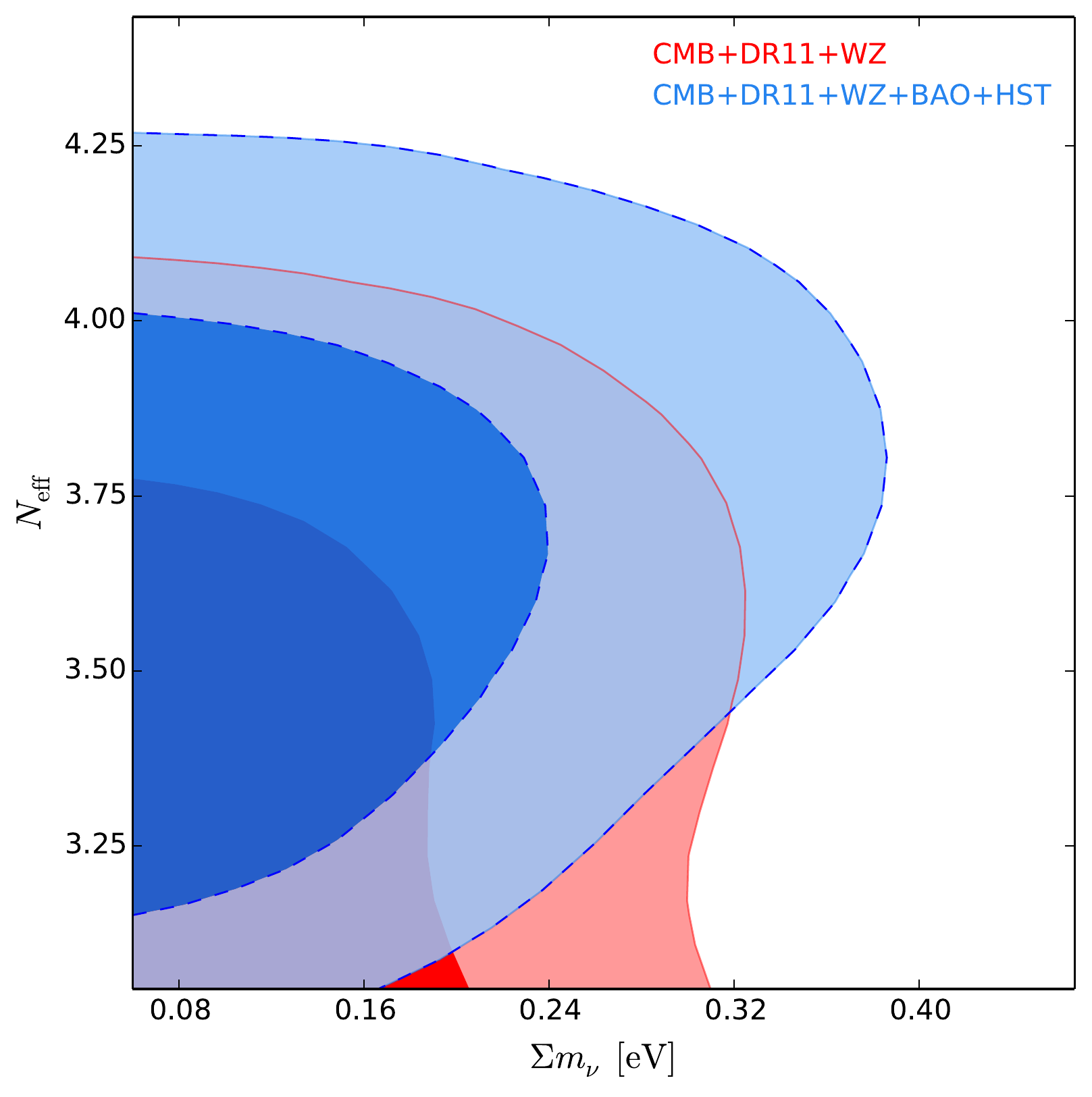}&\includegraphics[width=8.2cm]{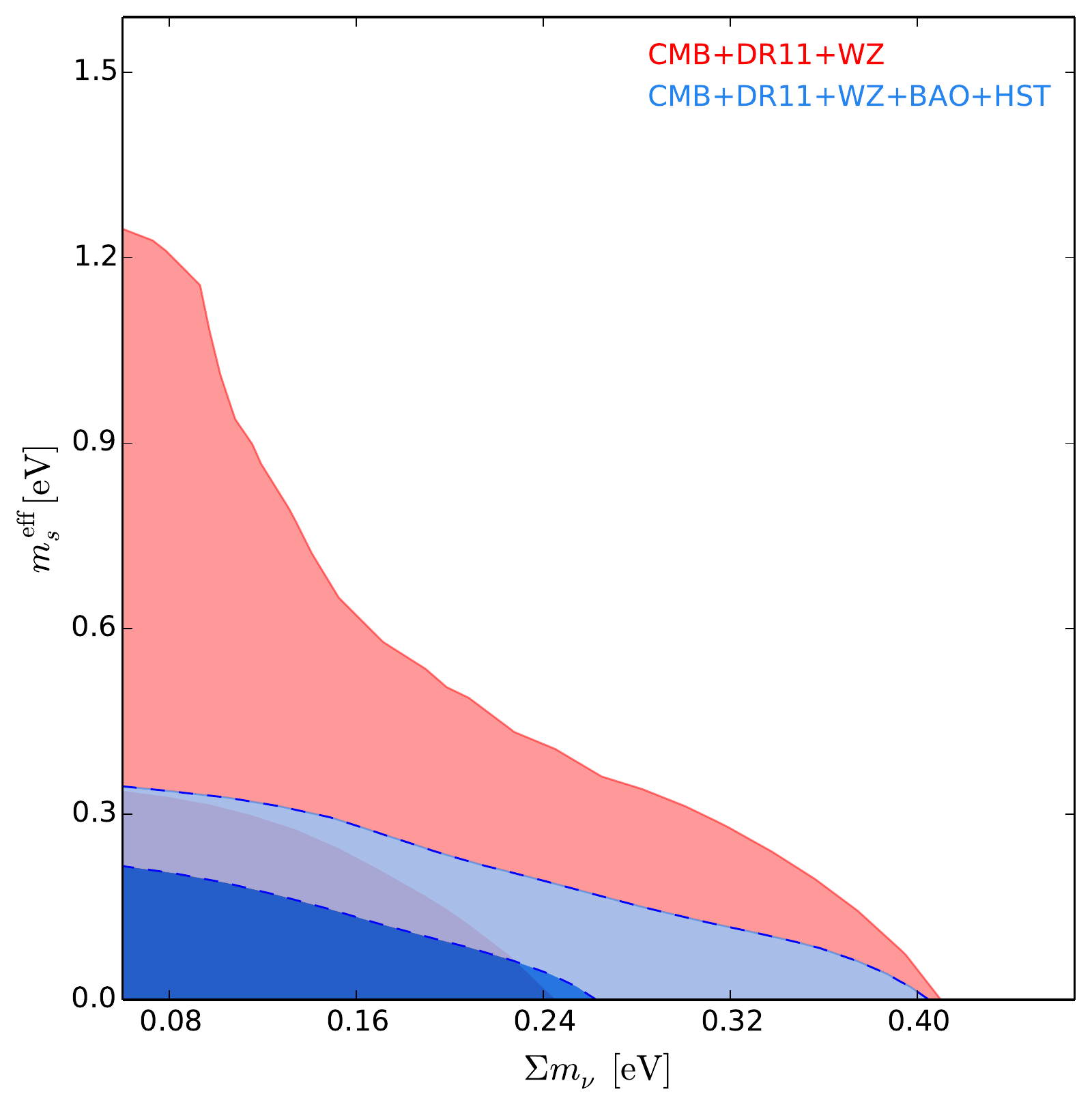}\\
\end{tabular}
 \caption{Left panel: the red contours show the $68\%$ and $95\%$~CL allowed
  regions from the combination of CMB data, BOSS DR11 BAO measurements
  and WiggleZ full shape power spectrum measurements in the ($\sum
m_\nu$ (eV), $\neff$) plane. The blue contours depict the
  constraints after a prior on the Hubble constant from
HST and the remaining BAO data are added in the analysis. Right panel: as in the left panel but in the ($\sum
m_\nu$ (eV), $m^\textrm{eff}_s$ (eV)) plane. }
\label{fig:meff}
\end{figure*}

Figure~\ref{fig:meff}, left panel, shows the degeneracy between 
the $\sum m_\nu$ and the total number of neutrino species $\neff$ 
(which accounts for the contribution of the three active neutrino
species plus $\Delta \neff$ massive sterile neutrinos). The red
contours depict the $68\%$ and $95\%$~CL allowed regions resulting
from the combination of CMB, BOSS DR11 BAO measurements, and full
shape power spectrum measurements from the WiggleZ survey. Notice that the allowed values of $\neff$ are
slightly larger than in the massless dark radiation scenario, since
sub-eV massive sterile neutrinos are contributing to the matter energy
density at the recombination period and therefore a larger  value of
$\neff$ will be required to leave unchanged both the angular location and
the height of the first acoustic peak.  The blue region depicts the results considering both the HST
$H_0$ prior and the remaining BAO data as well in the analysis. The right panel of  Fig.~\ref{fig:meff}, illustrates the degeneracy between the active and the sterile neutrino
masses, since both active and sterile sub-eV massive neutrinos contribute to the matter energy density at
decoupling, and both are free streaming relics which suppress
structure formation at small scales, after they become non relativistic.

\section{CMB constraints including the recent results from the BICEP2 experiment}

\begin{figure*}
\begin{tabular}{c c}
\includegraphics[width=8cm]{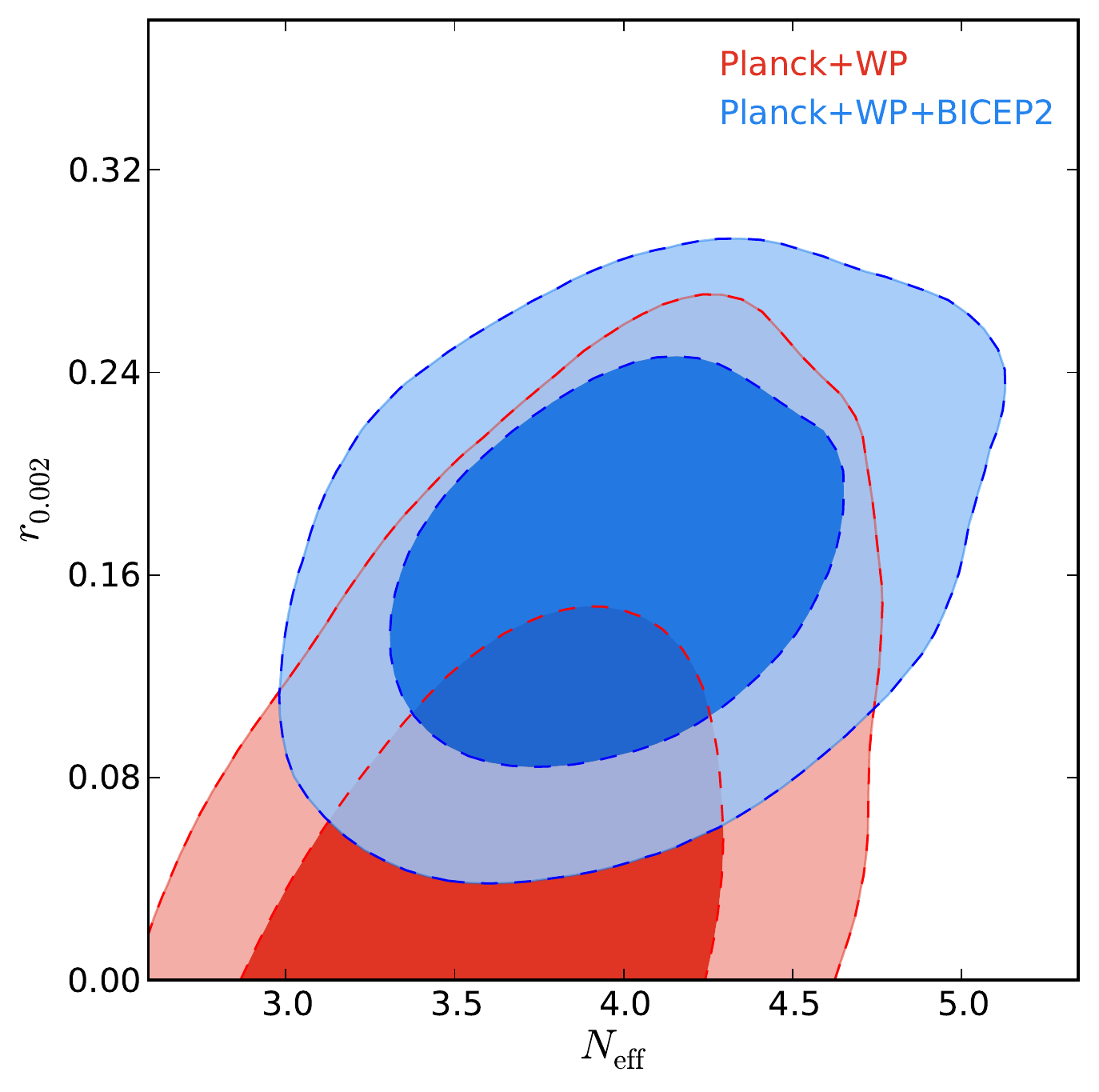}&\includegraphics[width=8.cm]{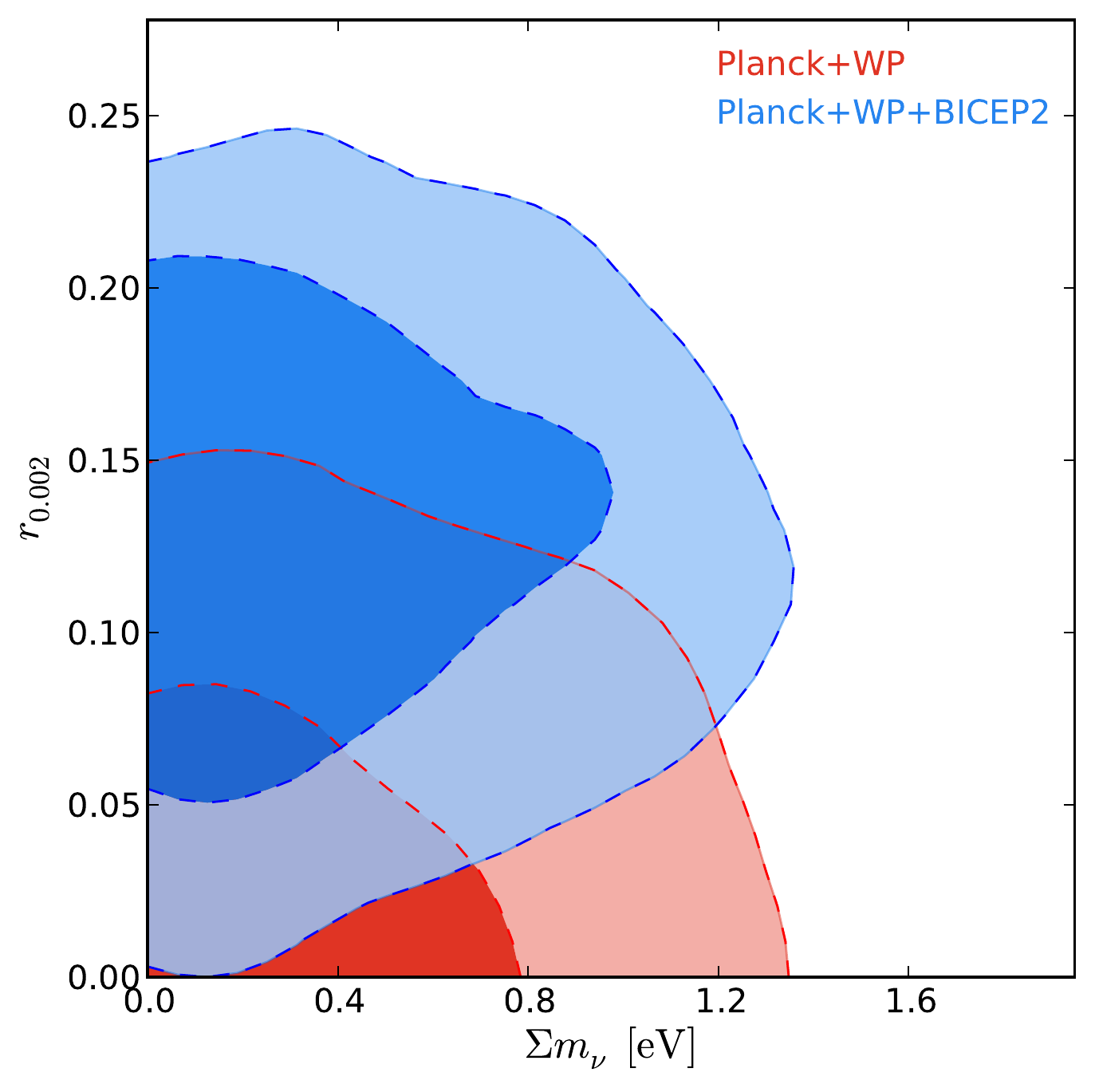}\\
\end{tabular}
 \caption{Left panel: Constraints in the $N_{eff}$ vs $r$ plane from
 Planck+WP and Planck+WP+BICEP2 data. Notice how the inclusion of
 the BICEP2 constraint shifts the contours towards $\neff>3$.
  Right panel: constraints on the $\Sigma m_{\nu}$ vs $r$ plane from Planck+WP
  and Planck+WP+BICEP2 data. In this case there is no indication for 
  neutrino masses from the combination of CMB data.}
\label{fig:bicep2}
\end{figure*}

Very recently, the BICEP2 experiment \cite{bicep2} claimed a detection at about $5.9\sigma$ for 
 B-mode polarization on large angular scales, compatible with the presence of
 a tensor component with amplitude $r_{0.002}=0.2_{-0.05}^{+0.06}$ at $68 \%$ 
 c.l.. It is therefore interesting to evaluate the impact of this measurement
for the effective number of relativistic species and neutrino masses.
We have therefore performed an analysis including a tensor component
(with zero running). The results are presented in Fig. \ref{fig:bicep2}.
As we can see, when the BICEP2 data are included, an extra background of
relativistic particle is preferred with $N_{eff}=4.00\pm0.41$ at $68 \%$ c.l..
CMB data alone is therefore suggesting a value for $\neff>3$ at good significance.
This result comes from the apparent tension between the Planck+WP limit of
$r<0.11$ at $95 \%$ c.l. and the recent BICEP2 result. This tension appears as
less evident when extra relativistic particles are included. We imagine a
further preference for $\neff>3$ if the HST data is included.
The BICEP2 result does not affect the current constraints on neutrino masses
as we can see from the right side of figure Fig. \ref{fig:bicep2}.

\section{Conclusions}
\label{sec:concl}
  
Standard cosmology includes hot thermal relics which refer to the three
light, active neutrino flavours of the Standard Model of elementary
particles. The largest effect of neutrino masses on the different
cosmological observables arises from their free streamig nature: the 
non-relativistic neutrino overdensities will contribute to clustering
only at scales larger than their
free streaming scale, suppressing the growth of matter density
fluctuations at small scales. CMB measurements from the Planck
satellite, including the lensing likelihood, low-$\ell$ polarization measurements from WMAP
9-year data and Baryon Acoustic Oscillation (BAO) measurements from a
number of surveys lead to the bound $\sum m_\nu<0.26$~eV at $95\%$~CL.

However, the existence of extra hot relic components, as dark
radiation relics, sterile neutrino species and/or thermal
axions will change the cosmological neutrino mass constraints. 
Dark radiation (i.e. purely massless species) may arise in several extentions of the
Standard Model of elementary particles, as, for instance, in
asymmetric dark matter models. On the other hand, the existence of extra massive species is well motivated by either 
the so-called neutrino oscillation anomalies (in the case of sterile
neutrino species) or by the strong CP problem (in the case of thermal
axions). Both extra, sterile neutrino species and axions have an associated free
streaming scale, reducing the growth of matter fluctuations at small
scales. These extra species will also contribute to the effective
number of relativistic degrees of freedom $\neff$, being 
$\neff=3.046$ the standard value, corresponding to the three active
neutrino contribution. The existence of extra light species at the Big Bang
Nucleosynthesis (BBN) epoch modifies the light element abundances,
especially the primordial helium mass fraction. 

We have presented here the constraints on the masses of the different thermal
relics in different scenarios using the available cosmological data in the beginning of this year 2014.
The data combination used here includes also the recent and most
precise distance BAO constraints to date from the BOSS Data Release 11
(DR11) results~\cite{Anderson:2013vga}, see also Refs.~\cite{Samushia:2013yga,Sanchez:2013tga,Chuang:2013wga}.
The tightest limit we find in the minimal three active massive
neutrino scenario is $\sum m_\nu < 0.22$~eV at $95\%$~CL from the
combination of CMB data, BAO data and HST measurements of the Hubble constant. 
 The addition of the constraints on $\sigma_8$ and $\Omega_m$ from the 
CFHTLens survey displaces the bounds on the neutrino mass to higher
values. However, the constraint on $\sigma_8$ and $\Omega_m$
from the Planck-SZ cluster catalog on galaxy number counts favours a non zero
value for the sum of the three active neutrino masses of $\sim 0.3$~eV
at $4\sigma$, see also Refs.~\cite{Hamann:2013iba,Wyman:2013lza}.
 
When considering simultaneously thermal axions and active massive
neutrino species, and including CMB, BOSS BAO DR11, additional BAO
measurements, WiggleZ power spectrum (full shape) information, the
$H_0$ HST prior and BBN light element abundances, the $95\%$~CL bounds
are $\sum m_\nu <0.25$~eV and $m_a<0.57$~eV  ($\sum m_\nu <0.21$~eV and $m_a<0.61$~eV) using recent (previous) deuterium estimates from
\cite{Cooke:2013cba} (\cite{fabio}) and helium  constraints from Ref.~\cite{Izotov:2013waa}.

Neither the addition of weak lensing
constraints on the $\sigma_8-\Omega_m$ relationship from the CFHTLens
experiment nor from  the Planck SZ cluster number counts favours non-zero thermal relic masses, except
for few cases in which the Planck SZ cluster number counts
information is considered together with the HST $H_0$ prior (or SNIa
luminosity distances) and all the BAO measurements. Only in this case there exists a mild  $\sim
2.2\sigma$ preference for a non zero axion mass of $0.6$~eV. Concerning neutrino
masses, there exists evidence for a neutrino mass of $\sim 0.2$~eV 
at the $\sim 3\sigma$ level exclusively for the case in which CMB data
is combined with BOSS BAO DR11 measurements and full-shape power spectrum information from the
WiggleZ  galaxy survey.

In the case in which we consider both massive
neutrinos and $\Delta \neff$ dark radiation species, the neutrino mass bounds are less stringent 
than in standard three neutrino massive case due to the 
large degeneracy between $\sum m_\nu$ and $\neff$, finding $\sum m_\nu < 0.31$~eV and
$\neff=3.45_{-0.54}^{+0.59}$ at $95\%$~CL from the combination of CMB
data and BOSS DR11 BAO measurements.  Contrarily to the massless
dark radiation case, but similarly to the thermal axion scenario, the addition of the
constraints on the $\sigma_8$ and $\Omega_m$ cosmological parameters
from the Planck-SZ cluster catalog on galaxy number counts does not
lead to a non zero value for the neutrino masses. After considering
the inclusion of Planck SZ clusters and
CFHTLens information to CMB data,
BOSS DR11 BAO, additional BAO measurements and the HST $H_0$ prior, 
the $95\%$~CL bounds on the active and the sterile neutrino parameters are $\sum m_\nu < 0.39$~eV,  $m^\textrm{eff}_s<0.59$~eV and
$\neff<4.01$. 

Big Bang Nucleosynthesis
  constraints  reduce both the mean value and the errors of
$\neff$ significantly. After the addition of the most recent measurements  of
deuterium~\cite{Cooke:2013cba} and helium~\cite{Izotov:2013waa}, and
using the theoretically derived fitting functions of Ref.~\cite{fabio}, we
find $\sum m_\nu < 0.24$~eV and $\neff=3.25_{-0.24}^{+0.25}$ at $95\%$~CL from the analysis of  CMB data,
WiggleZ power spectrum measurements and the HST $H_0$ prior finding no evidence for
$\neff>3$. If previous estimates of the deuterium primordial aundances are
used in the analysis~\cite{fabio},  there exists a $4 (2.5)\sigma$
preference for $\neff>3$,  with (without) HST data included in the
numerical analyses. If the additional sterile neutrino states are considered as massive
species, a $\sim 3.5 \sigma$ preference for $\neff>3$ still appears when
considering BBN measurements (with previous estimates of the deuterium
abundances from Ref.~\cite{fabio}) and the HST prior on the Hubble
constant.  The  $2.5-4\sigma$ preference for $\neff> 3$ always appears
for both the massless and the massive extra hot relic scenarios when
considering the theoretical fitting functions of Refs.~\cite{Steigman:2012ve,Cooke:2013cba},
independently of the deuterium measurements used in the analyses.
Accurate measurements as well as sharp theoretical predictions of the primordial deuterium
and helium light element abundances  are therefore crucial 
to constrain the value of $\neff$.

Finally, we have considered the recent B-mode polarization measurements
made by the BICEP2 experiment. Assuming that this detection is produced by
a primordial tensor component, we have found that in a LCDM$+r$ scenario
the presence of extra relativistic particles is significantly suggested by
current Planck+WP+BICEP2 data with $N_{eff}=4.00\pm0.41$ at $68 \%$ c.l..
An extra relativistic component therefore solves the current tension between
the Planck and BICEP2 experiments on the amplitude of tensor modes.

\section{Acknowledgments}
M.L. is supported by Ministero dell'Istruzione, dell'Universit\`a e
della Ricerca (MIUR) through the PRIN grant \emph{Galactic and extragalactic polarized microwave
emission'} (contract number PRIN 2009XZ54H2-002). Most of this work was carried out
while M.L. was visiting the Instituto de F\'isica Corpuscular, whose hospitality is
kindly acknowledged, supported by the grant \emph{Giovani ricercatori} of the University of Ferrara,
financed through the funds \emph{Fondi 5x1000 Anno 2010} and \emph{Fondi Unicredit
2013}. O.M. is supported by the Consolider Ingenio project CSD2007-00060, by
PROMETEO/2009/116, by the Spanish Ministry Science project FPA2011-29678 and by the ITN Invisibles PITN-GA-2011-289442.

\section{Appendix}
\label{sec:appdn}
For axion thermalization purposes, only the axion-pion interaction will be relevant.
To compute the axion decoupling temperature $T_D$ we follow the usual freeze out condition
\bea
\Gamma (T_D) = H (T_D)~.
\label{eq:freezeout}
\eea 
The average rate $\pi + \pi \rightarrow \pi
+a$ is given by~\cite{chang}:
\bea
\Gamma = \frac{3}{1024\pi^5}\frac{1}{f_a^2f_{\pi}^2}C_{a\pi}^2 I~,
\eea
where 
\bea
C_{a\pi} = \frac{1-R}{3(1+R)}~,
 \eea
is the axion-pion coupling constant~\cite{chang}, and 
\bea
I &=&n_a^{-1}T^8\int dx_1dx_2\frac{x_1^2x_2^2}{y_1y_2}
f(y_1)f(y_2) \nonumber \\
&\times&\int^{1}_{-1}
d\omega\frac{(s-m_{\pi}^2)^3(5s-2m_{\pi}^2)}{s^2T^4}~,
\eea
where  $n_a=(\zeta_{3}/\pi^2) T^3$ is the number density for axions in
thermal equilibrium, $f(y)=1/(e^y-1)$ denotes the pion distribution
function, 
$x_i=|\vec{p}_i|/T$,  $y_i=E_i/T$ ($i=1,2$), $s=2(m_{\pi}^2+T^2(y_1y_2-x_1x_2\omega))$, and we assume a common mass for the charged and neutral pions, $m_\pi=138$ MeV. 

We have numerically solved  the freeze out equation
Eq.~(\ref{eq:freezeout}), obtaining the axion decoupling temperature
$T_D$ versus the axion mass $m_a$ (or, equivalently, versus the axion
decay constant $f_a$). 
From the axion decoupling temperature, we can compute the current axion number density, related to the present photon density $n_\gamma=410.5 \pm 0.5$ cm$^{-3}$ via 
\bea
n_a=\frac{g_{\star S}(T_0)}{g_{\star S}(T_D)} \times \frac{n_\gamma}{2}~, 
\label{eq:numberdens}
\eea  
where $g_{\star S}$ refers to the number of \emph{entropic} degrees of
freedom. At the current temperature, $g_{\star S}(T_0) = 3.91$.





\begin{thebibliography}{99}

\frenchspacing
\bibitem{sergio}
J.~Lesgourgues and S.~Pastor,
  Adv.\ High Energy Phys.\  {\bf 2012}, 608515 (2012)
  [arXiv:1212.6154 [hep-ph]].
\bibitem{sergio2}
J.~Lesgourgues and S.~Pastor,
  Phys.\ Rept.\  {\bf 429}, 307 (2006)
  [astro-ph/0603494].
\bibitem{planck}
 P.~A.~R.~Ade {\it et al.}  [Planck Collaboration],
  arXiv:1303.5076 [astro-ph.CO].
\bibitem{lensingnu}
 J.~Lesgourgues, L.~Perotto, S.~Pastor and M.~Piat,
  Phys.\ Rev.\ D {\bf 73}, 045021 (2006)
  [astro-ph/0511735].
\bibitem{Reid:2009nq}
  B.~A.~Reid, L.~Verde, R.~Jimenez and O.~Mena,
  JCAP {\bf 1001}, 003 (2010)
  [arXiv:0910.0008 [astro-ph.CO]].
\bibitem{Hamann:2010pw}
  J.~Hamann, S.~Hannestad, J.~Lesgourgues, C.~Rampf and Y.~Y.~Y.~Wong,
  JCAP {\bf 1007}, 022 (2010)
  [arXiv:1003.3999 [astro-ph.CO]].
\bibitem{dePutter:2012sh} 
  R.~de Putter, O.~Mena, E.~Giusarma, S.~Ho, A.~Cuesta, H.~-J.~Seo, A.~J.~Ross and M.~White {\it et al.},
  Astrophys.\ J.\  {\bf 761}, 12 (2012)
  [arXiv:1201.1909 [astro-ph.CO]].
\bibitem{Giusarma:2012ph} 
  E.~Giusarma, R.~de Putter and O.~Mena,
  Phys.\ Rev.\ D {\bf 87}, no. 4, 043515 (2013)
  [arXiv:1211.2154 [astro-ph.CO]].
 \bibitem{Zhao:2012xw} 
  G.~-B.~Zhao, S.~Saito, W.~J.~Percival, A.~J.~Ross, F.~Montesano, M.~Viel, D.~P.~Schneider and M.~Manera {\it et al.},
  arXiv:1211.3741 [astro-ph.CO].
\bibitem{Hinshaw:2012fq} 
  G.~Hinshaw, D.~Larson, E.~Komatsu, D.~N.~Spergel, C.~L.~Bennett, J.~Dunkley, M.~R.~Nolta and M.~Halpern {\it et al.},
  arXiv:1212.5226 [astro-ph.CO].
\bibitem{Hou:2012xq} 
  Z.~Hou, C.~L.~Reichardt, K.~T.~Story, B.~Follin, R.~Keisler, K.~A.~Aird, B.~A.~Benson and L.~E.~Bleem {\it et al.},
  arXiv:1212.6267 [astro-ph.CO].
\bibitem{Sievers:2013wk} 
  J.~L.~Sievers, R.~A.~Hlozek, M.~R.~Nolta, V.~Acquaviva, G.~E.~Addison, P.~A.~R.~Ade, P.~Aguirre and M.~Amiri {\it et al.},
  arXiv:1301.0824 [astro-ph.CO].
\bibitem{Archidiacono:2013lva} 
  M.~Archidiacono, E.~Giusarma, A.~Melchiorri and O.~Mena,
  Phys.\ Rev.\ D {\bf 87}, 103519 (2013)
  [arXiv:1303.0143 [astro-ph.CO]].
\bibitem{Giusarma:2013pmn} 
  E.~Giusarma, R.~de Putter, S.~Ho and O.~Mena,
  Phys.\ Rev.\ D {\bf 88}, 063515 (2013)
  [arXiv:1306.5544 [astro-ph.CO]].
\bibitem{Archidiacono:2013fha} 
  M.~Archidiacono, E.~Giusarma, S.~Hannestad and O.~Mena,
  arXiv:1307.0637 [astro-ph.CO].
\bibitem{Riemer-Sorensen:2013jsa} 
  S.~Riemer-Sørensen, D.~Parkinson and T.~M.~Davis,
  arXiv:1306.4153 [astro-ph.CO].
  \bibitem{Hu:2014qma} 
  J.~-W.~Hu, R.~-G.~Cai, Z.~-K.~Guo and B.~Hu,
  arXiv:1401.0717 [astro-ph.CO].
\bibitem{Bennett:2012fp} 
  C.~L.~Bennett, D.~Larson, J.~L.~Weiland, N.~Jarosik, G.~Hinshaw, N.~Odegard, K.~M.~Smith and R.~S.~Hill {\it et al.},
  arXiv:1212.5225 [astro-ph.CO].
\bibitem{Riess:2011yx} 
  A.~G.~Riess, L.~Macri, S.~Casertano, H.~Lampeitl, H.~C.~Ferguson, A.~V.~Filippenko, S.~W.~Jha and W.~Li {\it et al.},
  Astrophys.\ J.\  {\bf 730}, 119 (2011)
  [Erratum-ibid.\  {\bf 732}, 129 (2011)]
  [arXiv:1103.2976 [astro-ph.CO]].
\bibitem{dr71}
 W.~J.~Percival {\it et al.}  [SDSS Collaboration],
  Mon.\ Not.\ Roy.\ Astron.\ Soc.\  {\bf 401}, 2148 (2010)
  [arXiv:0907.1660 [astro-ph.CO]].
\bibitem{dr72}
N.~Padmanabhan, X.~Xu, D.~J.~Eisenstein, R.~Scalzo, A.~J.~Cuesta, K.~T.~Mehta and E.~Kazin,
  arXiv:1202.0090 [astro-ph.CO].
\bibitem{wigglez}
C.~Blake, E.~Kazin, F.~Beutler, T.~Davis, D.~Parkinson, S.~Brough, M.~Colless and C.~Contreras {\it et al.},
  Mon.\ Not.\ Roy.\ Astron.\ Soc.\  {\bf 418}, 1707 (2011)
  [arXiv:1108.2635 [astro-ph.CO]].

\bibitem{Dawson:2012va} 
  K.~S.~Dawson {\it et al.}  [BOSS Collaboration],
  arXiv:1208.0022 [astro-ph.CO].
\bibitem{Eisenstein:2011sa} 
  D.~J.~Eisenstein {\it et al.}  [SDSS Collaboration],
  Astron.\ J.\  {\bf 142}, 72 (2011)
  [arXiv:1101.1529 [astro-ph.IM]].


\bibitem{anderson}
 L.~Anderson, E.~Aubourg, S.~Bailey, D.~Bizyaev, M.~Blanton, A.~S.~Bolton, J.~Brinkmann and J.~R.~Brownstein {\it et al.},
  Mon.\ Not.\ Roy.\ Astron.\ Soc.\  {\bf 427}, no. 4, 3435 (2013)
  [arXiv:1203.6594 [astro-ph.CO]].
\bibitem{6df}
 F.~Beutler, C.~Blake, M.~Colless, D.~H.~Jones, L.~Staveley-Smith, L.~Campbell, Q.~Parker and W.~Saunders {\it et al.},
  Mon.\ Not.\ Roy.\ Astron.\ Soc.\  {\bf 416}, 3017 (2011)
  [arXiv:1106.3366 [astro-ph.CO]].

\bibitem{dePutter:2014hza} 
  R.~de Putter, E.~V.~Linder and A.~Mishra,
  arXiv:1401.7022 [astro-ph.CO].

\bibitem{Hamann:2010bk} 
  J.~Hamann, S.~Hannestad, G.~G.~Raffelt, I.~Tamborra and Y.~Y.~Y.~Wong,
  Phys.\ Rev.\ Lett.\  {\bf 105}, 181301 (2010)
  [arXiv:1006.5276 [hep-ph]].
\bibitem{Giusarma:2011ex} 
  E.~Giusarma, M.~Corsi, M.~Archidiacono, R.~de Putter, A.~Melchiorri, O.~Mena and S.~Pandolfi,
  Phys.\ Rev.\ D {\bf 83}, 115023 (2011)
  [arXiv:1102.4774 [astro-ph.CO]].
\bibitem{Giusarma:2011zq} 
  E.~Giusarma, M.~Archidiacono, R.~de Putter, A.~Melchiorri and O.~Mena,
  Phys.\ Rev.\ D {\bf 85}, 083522 (2012)
  [arXiv:1112.4661 [astro-ph.CO]].
\bibitem{Hamann:2011ge} 
  J.~Hamann, S.~Hannestad, G.~G.~Raffelt and Y.~Y.~Y.~Wong,
  JCAP {\bf 1109}, 034 (2011)
  [arXiv:1108.4136 [astro-ph.CO]].
\bibitem{Melchiorri:2007cd} 
  A.~Melchiorri, O.~Mena and A.~Slosar,
  Phys.\ Rev.\ D {\bf 76}, 041303 (2007)
  [arXiv:0705.2695 [astro-ph]].
\bibitem{Hannestad:2007dd} 
  S.~Hannestad, A.~Mirizzi, G.~G.~Raffelt and Y.~Y.~Y.~Wong,
  JCAP {\bf 0708}, 015 (2007)
  [arXiv:0706.4198 [astro-ph]].
\bibitem{Hannestad:2008js} 
  S.~Hannestad, A.~Mirizzi, G.~G.~Raffelt and Y.~Y.~Y.~Wong,
  JCAP {\bf 0804}, 019 (2008)
  [arXiv:0803.1585 [astro-ph]].
\bibitem{Hannestad:2010yi} 
  S.~Hannestad, A.~Mirizzi, G.~G.~Raffelt and Y.~Y.~Y.~Wong,
  JCAP {\bf 1008}, 001 (2010)
  [arXiv:1004.0695 [astro-ph.CO]].
\bibitem{Archidiacono:2013cha} 
  M.~Archidiacono, S.~Hannestad, A.~Mirizzi, G.~Raffelt and Y.~Y.~Y.~Wong,
  JCAP {\bf 1310}, 020 (2013)
  [arXiv:1307.0615 [astro-ph.CO]].
\bibitem{Blennow:2012de} 
  M.~Blennow, E.~Fernandez-Martinez, O.~Mena, J.~Redondo and P.~Serra,
  JCAP {\bf 1207}, 022 (2012)
  [arXiv:1203.5803 [hep-ph]].
\bibitem{Diamanti:2012tg} 
  R.~Diamanti, E.~Giusarma, O.~Mena, M.~Archidiacono and A.~Melchiorri,
  Phys.\ Rev.\ D {\bf 87}, no. 6, 063509 (2013)
  [arXiv:1212.6007 [astro-ph.CO]].
\bibitem{Franca:2013zxa} 
  U.~Franca, R.~A.~Lineros, J.~Palacio and S.~Pastor,
  Phys.\ Rev.\ D {\bf 87}, 123521 (2013)
  [arXiv:1303.1776 [astro-ph.CO]].
\bibitem{Abazajian:2012ys} 
  K.~N.~Abazajian, M.~A.~Acero, S.~K.~Agarwalla, A.~A.~Aguilar-Arevalo, C.~H.~Albright, S.~Antusch, C.~A.~Arguelles and A.~B.~Balantekin {\it et al.},
  arXiv:1204.5379 [hep-ph].
\bibitem{Kopp:2013vaa} 
  J.~Kopp, P.~A.~N.~Machado, M.~Maltoni and T.~Schwetz,
  JHEP {\bf 1305}, 050 (2013)
  [arXiv:1303.3011 [hep-ph]].

\bibitem{Melchiorri:2008gq} 
  A.~Melchiorri, O.~Mena, S.~Palomares-Ruiz, S.~Pascoli, A.~Slosar and M.~Sorel,
  JCAP {\bf 0901}, 036 (2009)
  [arXiv:0810.5133 [hep-ph]].
\bibitem{Archidiacono:2012ri} 
  M.~Archidiacono, N.~Fornengo, C.~Giunti and A.~Melchiorri,
  Phys.\ Rev.\ D {\bf 86}, 065028 (2012)
  [arXiv:1207.6515 [astro-ph.CO]].
\bibitem{Archidiacono:2013xxa} 
  M.~Archidiacono, N.~Fornengo, C.~Giunti, S.~Hannestad and A.~Melchiorri,
  arXiv:1302.6720 [astro-ph.CO].
\bibitem{Mirizzi:2013kva} 
  A.~Mirizzi, G.~Mangano, N.~Saviano, E.~Borriello, C.~Giunti, G.~Miele and O.~Pisanti,
  Phys.\ Lett.\ B {\bf 726}, 8 (2013)
  [arXiv:1303.5368 [astro-ph.CO]].
\bibitem{Valentino:2013wha} 
  E.~Di Valentino, A.~Melchiorri and O.~Mena,
  JCAP {\bf 1311}, 018 (2013)
  [arXiv:1304.5981].
\bibitem{PecceiQuinn}
R.~D.~Peccei and H.~R.~Quinn,
  Phys.\ Rev.\ Lett.\  {\bf 38}, 1440 (1977);
 R.~D.~Peccei and H.~R.~Quinn,
  Phys.\ Rev.\  D {\bf 16}, 1791 (1977).
\bibitem{Hamann:2013iba} 
  J.~Hamann and J.~Hasenkamp,
  JCAP {\bf 1310}, 044 (2013)
  [arXiv:1308.3255 [astro-ph.CO]].
\bibitem{Wyman:2013lza} 
  M.~Wyman, D.~H.~Rudd, R.~A.~Vanderveld and W.~Hu,
  arXiv:1307.7715 [astro-ph.CO].
\bibitem{Ade:2013lmv} 
  P.~A.~R.~Ade {\it et al.}  [Planck Collaboration],
  arXiv:1303.5080 [astro-ph.CO].
\bibitem{Heymans:2013fya} 
  C.~Heymans, E.~Grocutt, A.~Heavens, M.~Kilbinger, T.~D.~Kitching, F.~Simpson, J.~Benjamin and T.~Erben {\it et al.},
  arXiv:1303.1808 [astro-ph.CO].

\bibitem{Cooke:2013cba} 
  R.~Cooke, M.~Pettini, R.~A.~Jorgenson, M.~T.~Murphy and C.~C.~Steidel,
  arXiv:1308.3240 [astro-ph.CO].
\bibitem{Izotov:2013waa} 
  Y.~I.~Izotov, G.~Stasinska and N.~G.~Guseva,
  arXiv:1308.2100 [astro-ph.CO].
\bibitem{Anderson:2013vga} 
  L.~Anderson, E.~Aubourg, S.~Bailey, F.~Beutler, V.~Bhardwaj, M.~Blanton, A.~S.~Bolton and J.~Brinkmann {\it et al.},
  arXiv:1312.4877 [astro-ph.CO].
\bibitem{Samushia:2013yga} 
  L.~Samushia, B.~A.~Reid, M.~White, W.~J.~Percival, A.~J.~Cuesta, G.~-B.~Zhao, A.~J.~Ross and M.~Manera {\it et al.},
  arXiv:1312.4899 [astro-ph.CO].

\bibitem{Sanchez:2013tga} 
  A.~G.~Sanchez, F.~Montesano, E.~A.~Kazin, E.~Aubourg, F.~Beutler, J.~Brinkmann, J.~R.~Brownstein and A.~J.~Cuesta {\it et al.},
  arXiv:1312.4854 [astro-ph.CO].

\bibitem{Chuang:2013wga} 
  C.~-H.~Chuang, F.~Prada, F.~Beutler, D.~J.~Eisenstein, S.~Escoffier, S.~Ho, J.~-P.~Kneib and M.~Manera {\it et al.},
  arXiv:1312.4889 [astro-ph.CO].
\bibitem{camb}
  A.~Lewis, A.~Challinor and A.~Lasenby,
  Astrophys.\ J.\  {\bf 538}, 473 (2000)
  [arXiv:astro-ph/9911177].
%
\bibitem{Lewis:2002ah}
  A.~Lewis and S.~Bridle,
  Phys.\ Rev.\  D {\bf 66}, 103511 (2002)
  [arXiv:astro-ph/0205436].

\bibitem{Ade:2013ktc} 
  P.~A.~R.~Ade {\it et al.}  [Planck Collaboration],
  arXiv:1303.5062 [astro-ph.CO].
  
\bibitem{Planck:2013kta} 
  P.~A.~R.~Ade {\it et al.}  [Planck Collaboration],
  arXiv:1303.5075 [astro-ph.CO].
  
\bibitem{Ade:2013tyw} 
  P.~A.~R.~Ade {\it et al.}  [Planck Collaboration],
  arXiv:1303.5077 [astro-ph.CO].
  
\bibitem{Das:2013zf} 
  S.~Das, T.~Louis, M.~R.~Nolta, G.~E.~Addison, E.~S.~Battistelli, J R.~Bond, E.~Calabrese and D.~C.~M.~J.~Devlin {\it et al.},
  arXiv:1301.1037 [astro-ph.CO].
  
\bibitem{Reichardt:2011yv} 
  C.~L.~Reichardt, L.~Shaw, O.~Zahn, K.~A.~Aird, B.~A.~Benson, L.~E.~Bleem, J.~E.~Carlstrom and C.~L.~Chang {\it et al.},
  Astrophys.\ J.\  {\bf 755}, 70 (2012)
  [arXiv:1111.0932 [astro-ph.CO]].

\bibitem{Eisenstein:1997ik} 
  D.~J.~Eisenstein and W.~Hu,
  Astrophys.\ J.\  {\bf 496}, 605 (1998)
  [astro-ph/9709112].

\bibitem{Parkinson:2012vd} 
  D.~Parkinson, S.~Riemer-Sorensen, C.~Blake, G.~B.~Poole, T.~M.~Davis, S.~Brough, M.~Colless and C.~Contreras {\it et al.},
  Phys.\ Rev.\ D {\bf 86}, 103518 (2012)
  [arXiv:1210.2130 [astro-ph.CO]].

\bibitem{snls}
 A.~Conley, J.~Guy, M.~Sullivan, N.~Regnault, P.~Astier, C.~Balland, S.~Basa and R.~G.~Carlberg {\it et al.},
  Astrophys.\ J.\ Suppl.\  {\bf 192}, 1 (2011)
  [arXiv:1104.1443 [astro-ph.CO]].
\bibitem{fabio}
F.~Iocco, G.~Mangano, G.~Miele, O.~Pisanti and P.~D.~Serpico,
  Phys.\ Rept.\  {\bf 472}, 1 (2009)
  [arXiv:0809.0631 [astro-ph]].
\bibitem{parthenope}
O.~Pisanti, A.~Cirillo, S.~Esposito, F.~Iocco, G.~Mangano, G.~Miele and P.~D.~Serpico,
  Comput.\ Phys.\ Commun.\  {\bf 178}, 956 (2008)
  [arXiv:0705.0290 [astro-ph]].
\bibitem{alterbbn}
A.~Arbey,
  Comput.\ Phys.\ Commun.\  {\bf 183}, 1822 (2012)
  [arXiv:1106.1363 [astro-ph.CO]].
\bibitem{Hou:2011ec} 
  Z.~Hou, R.~Keisler, L.~Knox, M.~Millea and C.~Reichardt,
  Phys.\ Rev.\ D {\bf 87}, 083008 (2013)
  [arXiv:1104.2333 [astro-ph.CO]].
\bibitem{Steigman:2012ve} 
  G.~Steigman,
  Adv.\ High Energy Phys.\  {\bf 2012}, 268321 (2012)
  [arXiv:1208.0032 [hep-ph]].

\bibitem{chang}
  S.~Chang and K.~Choi,
  Phys.\ Lett.\  B {\bf 316}, 51 (1993)
  
\bibitem{bicep2}
P.~A.~R.~Ade {\it et al.}  [BICEP2 Collaboration],
  arXiv:1403.3985 [astro-ph.CO].

\end{thebibliography}
\end{document}